\documentclass[12pt]{article}
\usepackage{amsmath}
\usepackage{graphicx,psfrag,epsf}
\usepackage{enumerate}
\usepackage{natbib}
\usepackage{amsthm}
\usepackage{amssymb}
\usepackage[colorlinks,citecolor=blue,urlcolor=blue,filecolor=blue,backref=page]{hyperref}
\usepackage{bbm}
\usepackage{bm}
\usepackage{dcolumn}
\newcolumntype{d}[1]{D{.}{.}{#1}}

\usepackage{booktabs,siunitx}
\usepackage{multirow}

\usepackage{enumitem}

\usepackage{algorithm}
\usepackage{algpseudocode}

\newcommand{\Nd}{\mathcal{N}}

\newcommand{\B}{\mathbf{B}}

\newcommand\indep{\protect\mathpalette{\protect\independenT}{\perp}}
\def\independenT#1#2{\mathrel{\rlap{$#1#2$}\mkern2mu{#1#2}}}

\newtheorem{assumption}{Assumption}

\newcommand{\blind}{0}

\begin{document}

\def\spacingset#1{\renewcommand{\baselinestretch}%
{#1}\small\normalsize} \spacingset{1}


\newcommand{\tit}{\bf Bayesian Propensity Score-Augmented Latent Factor Models for Causal Inference 
with Time-Series Cross-Sectional Data}

\if0\blind
{
  \title{\tit}
  \author{Licheng Liu\thanks{
     Assistant Professor, Department of Political Science, University of Michigan.
     \\Email:
     \href{lichengl@umich.edu}{lichengl@umich.edu}.
     }
  }
    
  \maketitle
} \fi

\if1\blind
{
   \title{\tit}

   \maketitle
} \fi

\bigskip
\begin{abstract}

We propose a Bayesian propensity score–augmented latent factor model for causal inference with 
time-series cross-sectional data. The framework explicitly models the treatment assignment mechanism 
by incorporating latent factor loadings, while the outcome model flexibly incorporates the propensity 
score, for example through stratification. Relative to existing approaches, the proposed method 
provides greater flexibility and captures additional heterogeneity across propensity-score strata, 
enabling more credible comparisons between treated and control units within each stratum. For 
estimation and inference, we adopt an approximate Bayesian procedure to address the model feedback 
problem common in Bayesian propensity score analysis. We demonstrate the performance of the proposed 
method through Monte Carlo simulations and an empirical application examining the effect of political 
connections on firm value.
\end{abstract}

\noindent%
\vfill

\newpage
\spacingset{1.45} 
\section{Introduction}

The propensity score–based approach \citep{rosenbaum1983central} has been widely adopted for 
estimating treatment effects from observational studies. Unlike experimental settings, where the 
treatment assignment mechanism is known to the researcher, observational studies typically rely on 
unconfoundedness assumptions to draw causal inferences because researchers do not directly control 
treatment assignment. One example is the \textit{conditional ignorability} assumption, which 
posits that treatment assignment is as-if randomized conditional on a set of observed pre-treatment 
confounders. The propensity score is therefore defined as the probability of receiving treatment 
given these pre-treatment confounders.

The propensity score possesses an important balancing property: conditional on the propensity score, 
treatment assignment is independent of the observed confounders. In other words, the propensity score 
summarizes the information in the confounders relevant to treatment assignment. This property enables 
causal estimation using methods such as matching, inverse probability weighting, and 
subclassification, which aim to balance the distribution of pre-treatment confounders between treated 
and control groups.

Despite these appealing properties, the use of propensity score methods in time-series cross-sectional 
(TSCS) settings has developed more slowly. Most existing approaches continue to rely on variants of 
the conditional ignorability assumption when modeling the probability that a unit belongs to the 
treated group, both in canonical difference-in-differences settings 
(e.g., \citealp{abadie2005semiparametric}) and in staggered adoption designs 
(e.g., \citealp{callaway2021difference}). In many cases, the propensity score is combined with an 
outcome regression model to achieve \emph{double robustness}, whereby consistent estimation of 
treatment effects is obtained if either the treatment assignment model or the outcome model is 
correctly specified.

A defining feature of causal inference with TSCS data, however, is the likely presence of unobserved 
confounders. Model-based approaches typically approximate such confounders using additive two-way 
fixed effects or interactive fixed-effects structures (e.g., 
\citealp{abadie2010synthetic,xu2017generalized}). The associated identification assumption is often 
referred to as \textit{latent ignorability} \citep{frangakis1999addressing}, which relaxes standard 
conditional ignorability by allowing for unobserved confounding. Nevertheless, many existing 
model-based panel methods (e.g., \citealp{athey2021matrix}) do not explicitly specify the treatment 
assignment mechanism or derive a corresponding propensity score under latent ignorability. This is 
largely because latent confounders themselves must be estimated before a propensity score can be 
constructed. Only recently have methods been proposed that explicitly model treatment assignment 
as depending on latent confounders in panel data settings 
(e.g., \citealp{forino2025propensity,schmidt2025bayesian}).

To address these challenges, we propose a novel Bayesian framework that jointly models treatment 
assignment and potential outcomes under control as functions of both observed pre-treatment 
covariates and latent confounders, enabling treatment effect estimation in time-series cross-sectional 
data. Specifically, we introduce a Bayesian propensity score–augmented latent factor model (PS-LFM), in 
which treatment assignment depends explicitly on latent confounders while accounting for uncertainty in 
the estimation of both latent confounders and propensity scores.

Compared with conventional frequentist propensity score approaches, the proposed Bayesian framework 
formally incorporates uncertainty in propensity score estimation into treatment effect inference 
\citep{mccandless2009bayesian,alvarez2021uncertain}. We further establish identification of treatment 
effects under latent ignorability and demonstrate the necessity of explicitly modeling the treatment 
assignment mechanism in the presence of unobserved confounding.

Because latent confounders enter both the treatment assignment and outcome models, and because the 
propensity score also appears in the outcome model\footnote{This corresponds to the outcome stage of 
propensity score analysis in \citet{rosenbaum1983central}.}, a fully Bayesian analysis may induce 
substantial model feedback problem \citep{zigler2013model}. Such feedback can lead to inaccurate 
counterfactual predictions and biased treatment effect estimates. To address this issue, we adopt an 
\textit{approximate Bayesian} analysis \citep{zigler2013model,zigler2014uncertainty} for counterfactual 
prediction and treatment effect estimation, an approach shown to effectively mitigate model feedback 
problems in Bayesian propensity score analysis. The proposed PS-LFM is implemented and publicly 
available in the open-source R package \texttt{bpCausal}.

This paper contributes to the existing literature in three main ways. First, it contributes to the 
design-based causal inference literature for TSCS data (e.g., 
\citealp{abadie2005semiparametric,callaway2021difference,arkhangelsky2022doubly}) by explicitly 
modeling the treatment assignment mechanism under latent ignorability. Second, it contributes to 
the Bayesian causal inference literature \citep{li2023bayesian} by 
extending Bayesian 
propensity score analysis, previously developed primarily for cross-sectional settings (e.g., 
\citealp{mccandless2009bayesian,zigler2013model,zigler2014uncertainty,alvarez2021uncertain}), to TSCS 
data with potentially rich unobserved confounding structures. Finally, it contributes to the panel 
data modeling literature with grouped structures and interactive fixed effects 
(e.g., \citealp{mehrabani2025shrinkage}) by allowing the latent grouping structure, 
through subclassification, to depend on the propensity score.

The remainder of the paper is organized as follows. Section 2 introduces the general framework, 
establishes identification results under the stated assumptions, and presents the proposed methodology. 
Section 3 describes the Bayesian estimation and inference for the proposed PS-LFM. Section 4 presents 
Monte Carlo studies, and Section 5 illustrates the applicability of the method through a reanalysis of 
the effect of political connections on firm value. The final section concludes with a discussion of the 
method's limitations.

\section{Methodology}

\subsection{Setup and notation}

We consider a panel dataset consisting of $N$ units observed over $T$ time periods, indexed by 
$i = 1, 2, \ldots, N$ and $t = 1, 2, \ldots, T$, respectively. For each unit--time pair, we observe 
an outcome variable $Y_{it} \in \mathbb{R}$, for which the effect of a binary treatment 
$D_{it} \in \{0,1\}$ is the primary quantity of interest. For each unit $i$, we also observe a 
($p \times 1$) vector of pre-treatment covariates $Z_i = (Z_{i,1}, \ldots, Z_{i,p})$ 
\footnote{It is straightforward to include exogenous 
time-varying covariates $X_{it}$ in the proposed methodology. However, we omit the 
time-varying covariates and just consider time-invariant $Z_i$ to illustrate treatment 
assignment mechanism.}.

We focus on a setting with staggered adoption of the treatment. Let $1 < T_0 < T$ denote the first 
period in which any unit receives treatment. Define $A_i$ as the adoption time for unit $i$, and 
set $A_i = T + 1$ if unit $i$ never adopts the treatment during the observation window. The 
treatment path is therefore fully determined by the adoption date, as 
$
D_{it} = \mathbbm{1}\{t \ge A_i\}
$ for $ t = 1, \ldots, T$.
Let $\bm{D}_i = (D_{i1}, \ldots, D_{iT})$ denote the treatment sequence for unit $i$. 
We impose the following assumption on the assignment of adoption dates.

\begin{assumption}[Randomization of adoption time for treated units]\label{rand}
Conditional on $A_i \neq T + 1$, the adoption time $A_i$ is randomly drawn from 
$\{T_0, T_0+1, \ldots, T\}$ for each $i \in \{1, \ldots, N\}$.
\end{assumption}

Assumption~\ref{rand} states that, among units that eventually receive treatment, the timing of 
adoption is randomly assigned and does not depend on unit-level covariates. Accordingly, we define 
$i \in \mathcal{T}$ (treated units) if $A_i \in \{T_0, \ldots, T\}$ and 
$i \in \mathcal{C}$ (control units) if $A_i = \infty$. With a slight abuse of notation, let 
$W_i = \mathbbm{1}\{i \in \mathcal{T}\}$ denote an indicator for whether unit $i$ ever receives 
treatment. Under this assumption, unit-level covariates may influence whether a unit is ever treated 
(i.e., $W_i$), but conditional on $W_i=1$, the adoption timing $A_i$ is random. This assumption is 
satisfied, for example, when treatment timing is externally randomized, but it may be violated if 
unit-level characteristics systematically affect adoption timing, as in the setting of 
\citet{callaway2021difference}.

We adopt the potential outcomes framework \citep{rubin1974estimating} to define treatment effects. 
The potential outcome for unit $i$ at time $t$ is denoted $Y_{it}(\bm{d}_i)$, where 
$\bm{d}_i = (d_1, \ldots, d_T) \in \{0, 1\}^T $ represents a realization of the full treatment 
sequence. Following \citet{arkhangelsky2022doubly} and \citet{athey2022design}, We impose the 
following assumption of exclusion restriction, which rules out effects of both past and future 
treatments.

\begin{assumption}[Exclusion restriction]\label{er}
There is no anticipation of future treatment, and no carryover effects from past treatment. 
Consequently,
\[
Y_{it}(\bm{d}_i) = Y_{it}(d_{it}),
\]
for all $i$, $t$, and $d_{it} \in \{0,1\}$.
\end{assumption}

Under Assumption~\ref{er}, there are only two potential outcomes for each unit-time pair: 
$Y_{it}(1)$ and $Y_{it}(0)$. The individual treatment effect is therefore defined as:
\[
\delta_{it} = Y_{it}(1) - Y_{it}(0).
\]
Our primary estimand is the average treatment effect on the treated (ATT) $r$ periods after treatment 
adoption, defined as:
\begin{equation}
ATT_r =
\frac{\sum_{i,t} \mathbbm{1}\{A_i=t\}\,\delta_{i,t+r}}
{\sum_{i,t} \mathbbm{1}\{A_i=t\}}.
\end{equation}
This quantity is a convex combination of individual treatment effects and represents the dynamic 
effect relative to treatment adoption. We also consider the overall average treatment effect on 
the treated, defined as
$
ATT = \frac{\sum_{i,t} D_{it}\delta_{it}}{\sum_{i,t} D_{it}}
$, which captures the average effect for treated units over the entire sample period.

To identify the treatment effects above, we assume the presence of both observed 
pre-treatment confounders $Z_i$ and unobserved, time-invariant unit-specific confounders $\gamma_i$. 
We formalize this using a latent ignorability assumption \citep{frangakis1999addressing}:

\begin{assumption}[Latent ignorability]\label{asm:panelig}
\[
\{Y_{it}(0), Y_{it}(1)\} \indep W_i \mid Z_i, \gamma_i,
\]
for all $i$ and $t$.
\end{assumption}

This assumption states that the potential outcomes are independent of whether a unit is treated, 
conditional on $Z_i$ and $\gamma_i$. In addition, let $\Pr(W_i=1 \mid Z_i,\gamma_i)$ denote the 
propensity score for treatment. We in addition impose the positivity assumption, i.e.,
$
0 < \Pr(W_i=1 \mid Z_i,\gamma_i) < 1
$,
which ensures overlap in treatment assignment conditional on both observed and latent confounders. 

Relative to standard conditional ignorability assumptions that rule out unobserved confounding, 
Assumption~\ref{asm:panelig} explicitly allows for time-invariant latent confounders, making it 
particularly suitable for time-series cross-sectional settings.

\subsection{Bayesian causal inference}

We establish identification of the treatment effects within the Bayesian causal inference framework 
\citep{rubin1978bayesian}. The primary objective of Bayesian causal inference is to derive the 
posterior predictive distributions of missing potential outcomes. In our setting, because latent 
confounders are unobserved, we must jointly model the posterior distribution of the missing potential 
outcomes and the latent confounders, and then integrate over the posterior distribution of the latter.
Since our estimands are the average treatment effects on the treated (ATTs), we focus on imputing the 
missing counterfactual outcomes under control for treated observations.

Let 
$\bm{Y}_i(0) = \left(Y_{i1}(0), \ldots, Y_{iT}(0)\right)$ denote the trajectory of potential outcomes 
under control for unit $i$, and stack these to form 
$\bm{Y}(0) = \left(\bm{Y}_1(0)', \ldots, \bm{Y}_N(0)'\right)$ for all units. Likewise, let 
$\bm{W} = (W_1, \ldots, W_N)$ denote the treatment group indicators, 
$\bm{Z} = (Z_1, \ldots, Z_N)$ the observed covariates, and 
$\Gamma = (\gamma_1, \ldots, \gamma_N)$ the latent confounders. 
We follow \citet{rubin2005causal} and call 
$f\left(\bm{Y}(0), \bm{Z}, \Gamma\right)$ a model on the ``science'' 
\footnote{As in  \citet{rubin2005causal}, the science means the complete data 
$\left(\bm{Y}(0), \bm{Y}(1), \bm{Z}, \Gamma\right)$. Here we 
omit $\bm{Y}(1)$ to focus on the imputation of missing $\bm{Y}(0)$.}. 
Assuming the observations of each unit are exchangable conditional on $Z_i$ and $\gamma_i$,  
by de Finetti's theorem, we can write 
\begin{equation}\label{miss}
f(\bm{Y}(0), \bm{Z}, \Gamma)
= \int \prod_{i = 1}^{N} f(Z_i, \gamma_i, \bm{Y}_i(0) | \theta) f(\theta) d\theta,    
\end{equation}
where $\theta$ represents a vector of parameters in the potential outcome model 
$f(\cdot|\theta)$, with $f(\theta)$ its prior distribution.

We partition the potential outcomes under control as 
$\bm{Y}(0) = \left(\bm{Y}(0)^{obs}, \bm{Y}(0)^{mis}\right)$, where $\bm{Y}(0)^{mis}$ represents the 
missing counterfactual outcomes for treated observations. Because $\Gamma$ is unobserved, 
we derive the posterior distribution of $\bm{Y}(0)^{mis}$ conditional on 
$\bm{Y}(0)^{obs}$, $\bm{W}$, $\bm{Z}$ as follows:
\begin{equation}\label{id}
\begin{aligned}
&f(\bm{Y}(0)^{mis} \mid \bm{Y}(0)^{obs}, \bm{W}, \bm{Z}) \\
\propto\;& f(\bm{W}, \bm{Z}, \bm{Y}(0)) \\
=\;& \int f(\bm{W}, \bm{Z}, \Gamma, \bm{Y}(0)) d \Gamma  \\
=\;& \int f(\bm{W} \mid \bm{Z}, \Gamma, \bm{Y}(0)) f(\bm{Z}, \Gamma, \bm{Y}(0)) d \Gamma \\
=\;& \int f(\bm{W} \mid \bm{Z}, \Gamma)
f(\bm{Y}(0)^{mis} \mid \bm{Z}, \Gamma, \bm{Y}(0)^{obs})
f(\bm{Z}, \Gamma, \bm{Y}(0)^{obs}) d \Gamma \\
\propto\;& \int f(\bm{W} \mid \bm{Z}, \Gamma)
f(\bm{Y}(0)^{mis} \mid \bm{Z}, \Gamma, \bm{Y}(0)^{obs})
f(\Gamma \mid \bm{Z}, \bm{Y}(0)^{obs}) d \Gamma,
\end{aligned}
\end{equation}
where the fifth line of Equation~(\ref{id}) follows from Assumption~\ref{asm:panelig}, which 
implies independence between treatment assignment $\bm{W}$ and potential outcomes $\bm{Y}(0)$ 
conditional on $\bm{Z}$ and $\Gamma$. 

It is worth noting that the posterior predictive distribution~(\ref{id}) of $\bm{Y}(0)^{mis}$ can 
be expressed as $f(\bm{Y}(0)^{mis} \mid \bm{Z}, \Gamma, \bm{Y}(0)^{obs})$, the second component in 
the final integral, marginalized over the posterior distribution of the latent confounders $\Gamma$. 
This posterior is proportional to the product of the propensity of treatment 
$f(\bm{W} \mid \bm{Z}, \Gamma)$ and the conditional distribution 
$f(\Gamma \mid \bm{Z}, \bm{Y}(0)^{obs})$. The first term characterizes the treatment assignment 
mechanism, while the second reflects the extent to which $\Gamma$ can be learned from the observed data.

In the factor analysis literature, inference on latent factors such as $\Gamma$ typically relies on 
additional structural assumptions that enable their recovery from observed outcomes, sometimes 
referred to as ``flexible data extraction'' (e.g., \citealp{xu2017generalized,pang2022bayesian}). The 
robustness of treatment effect estimates to violations of these assumptions has been examined in panel 
data settings; see, for example, \citet{liu2025bayesian}.

According to Equation~(\ref{miss}), the posterior predictive distribution above 
can be further expressed as:
\begin{equation}\label{id2}
\begin{aligned}
&f(\bm{Y}(0)^{mis} \mid \bm{Y}(0)^{obs}, \bm{W}, \bm{Z}) \\
\propto\;& \int f(\bm{W} \mid \bm{Z}, \Gamma)
f(\bm{Y}(0)^{mis} \mid \bm{Z}, \Gamma, \bm{Y}(0)^{obs})
f(\Gamma \mid \bm{Z}, \bm{Y}(0)^{obs}) d \Gamma \\
=\;& \iint f(\bm{W} \mid \bm{Z}, \Gamma)
f(\bm{Y}(0)^{mis} \mid \bm{Z}, \Gamma, \theta) 
f(\theta \mid \bm{Z}, \Gamma, \bm{Y}(0)^{obs})
f(\Gamma \mid \bm{Z}, \bm{Y}(0)^{obs}) d \Gamma d\theta \\ 
=\;& \iint f(\bm{W} \mid \bm{Z}, \Gamma)
f(\bm{Y}(0)^{mis} \mid \bm{Z}, \Gamma, \theta) 
f(\theta, \Gamma \mid \bm{Z}, \bm{Y}(0)^{obs}) d \Gamma d\theta, 
\end{aligned}
\end{equation}
where $f(\theta \mid \bm{Z}, \Gamma, \bm{Y}(0)^{obs})$ in the third line of Equation~(\ref{id2}) 
denotes the posterior distribution of the model parameters $\theta$ conditional on 
$\bm{Z}$, $\Gamma$, and $\bm{Y}(0)^{obs}$.

Overall, the re-expressed posterior predictive distribution~(\ref{id2}) indicates that 
identification and estimation of the counterfactuals 
rely on three components: modeling the treatment assignment mechanism, 
modeling the potential outcomes under control, and learning both the model parameters and latent 
confounders from the observed data.

\subsection{Propensity score-augmented outcome model}

The latent ignorability Assumption~\ref{asm:panelig} implies that identification of treatment effect 
of $D_{it}$ on $Y_{it}$ requires controlling for both observed confounders $Z_i$ and unobserved 
time-invariant confounders $\gamma_i$. In model-based approaches to causal inference, the effect 
of $\gamma_i$ is often captured either through additive two-way fixed effects in 
difference-in-differences designs or through latent factor structures in interactive fixed-effects 
models (e.g., \citealp{bai2009panel}). A general functional form for the potential outcome model can 
be written as
\begin{equation}\label{ob_model}
\begin{gathered}
Y_{it}(0) = g(Z_i, \gamma_i, \varepsilon_{it}), \\
Y_{it}(1) = \delta_{it} D_{it} + Y_{it}(0),
\end{gathered}
\end{equation}
where $\delta_{it}$ denotes heterogeneous treatment effects and 
$\varepsilon_{it}$ is a mean-zero error term.

Note that the potential outcome model in Equation~(\ref{ob_model}) corresponds exactly to the 
second component in the final integral of the posterior~(\ref{id2}).
Given an outcome model, researchers can impute the counterfactual 
outcomes under control for treated observations and then estimate the average treatment effect on the 
treated by averaging the differences between observed outcomes and predicted counterfactuals. 
Such counterfactual prediction–based approaches are referred to as ``causal panel data models'' in 
\citet{athey2021matrix}. For example, a linear specification incorporating $\gamma_i$  
through a latent factor structure can be written as
\begin{equation}\label{model1}
Y_{it} = \delta_{it} D_{it} + Z_i' \alpha_t + \gamma_i' f_t + \varepsilon_{it},
\end{equation}
where $\alpha_t$ denotes time-varying coefficients for the observed pre-treatment covariates $Z_i$. 
Here, $\gamma_i$ can be interpreted as an $(r \times 1)$ vector of factor loadings and $f_t$ as an 
$(r \times 1)$ vector of common factors, with the number of factors $r$ typically unknown 
\textit{a priori}. This specification implies a latent factor model,
$
Y_{it}(0) = Z_i' \alpha_t + \gamma_i' f_t + \varepsilon_{it}
$,
for potential outcomes under control.  

Despite its popularity as an empirical estimation strategy, the causal latent factor 
model~(\ref{model1}) typically does not explicitly model the treatment assignment mechanism, relying 
instead on the latent ignorability assumption. One reason is that modeling treatment assignment 
requires accounting for latent confounders, which must themselves be inferred from observed data, 
as discussed above. Moreover, because the propensity score, defined as $Pr(W_i = 1 \mid Z_i, \Gamma_i)$, 
summarizes information contained in both $Z_i$ and the latent confounders $\gamma_i$, omitting 
this term from the outcome model may lead to model mis-specification and, consequently, biased estimates 
of treatment effects. Only recently have methods been developed that explicitly model treatment 
assignment under latent ignorability (e.g., \citealp{forino2025propensity,schmidt2025bayesian}).

To address this limitation, we extend the above framework by explicitly incorporating the treatment 
assignment mechanism as an additional stage. Specifically, we propose the following joint model for 
treatment assignment and potential outcomes:
\begin{equation}\label{dr}
\begin{gathered}
W_i = \mathbbm{1}\{Z_i' \lambda_z + \gamma_i' \lambda_\gamma + \nu_i \ge 0\}, \\
Y_{it}(0) =
g(Z_i, ps(Z_i,\gamma_i;\bm{\lambda}))' \beta_t
+ \gamma_i' f_t + \varepsilon_{it}, \\
Y_{it}(1) = \delta_{it} D_{it} + Y_{it}(0),
\end{gathered}
\end{equation}
where the first equation specifies treatment assignment as a function of both observed covariates 
$Z_i$ and latent factor loadings $\gamma_i$, allowing unobserved confounders to influence both 
treatment assignment and outcomes. This component corresponds to the propensity score stage in the 
Bayesian propensity score literature (e.g., \citealp[]{mccandless2009bayesian,zigler2013model}). 

The propensity score implied by the first equation in~(\ref{dr}), defined as
$ps(Z_i,\gamma_i;\bm{\lambda}) = \Pr(W_i=1 \mid Z_i,\gamma_i)$, enters the outcome model directly 
through a flexible but prespecified function $g(\cdot)$, together with $Z_i$ and $\gamma_i$. 
This component constitutes the outcome stage. We allow these effects to vary over time through the 
coefficient vector $\beta_t$. The latent factor term $\gamma_i' f_t$ serves as a residual adjustment, 
capturing variation not explained by the propensity score–based component. We refer to the proposed 
joint model~(\ref{dr}) as the propensity score–augmented latent factor model (PS-LFM).

For the propensity score stage, without loss of generality, we assume 
$\nu_i \sim \mathcal{N}(0,1)$, which implies
\begin{equation}
ps(Z_i,\gamma_i;\bm{\lambda})
= \Phi(Z_i' \lambda_z + \gamma_i' \lambda_\gamma),
\end{equation}
where $\Phi(\cdot)$ denotes the standard normal cumulative distribution function.

For the outcome model, the function $g(Z_i, ps(Z_i,\gamma_i;\bm{\lambda}))$ specifies how the 
propensity score enters the outcome model. One example is a propensity-score subclassification 
specification with $k$ strata \citep{rosenbaum1983central,mccandless2009bayesian}. For instance, 
when $k=3$ and $0<q_1 < q_2<1$ are two threshold values,
\begin{equation}\label{strata}
g(Z_i, ps(Z_i,\gamma_i;\bm{\lambda})) =
\begin{cases}
[1,0,0]' \otimes Z_i, & 0 < ps(Z_i,\gamma_i;\bm{\lambda}) < q_1, \\
[0,1,0]' \otimes Z_i, & q_1 \le ps(Z_i,\gamma_i;\bm{\lambda}) < q_2, \\
[0,0,1]' \otimes Z_i, & q_2 \le ps(Z_i,\gamma_i;\bm{\lambda}) < 1. \\
\end{cases}
\end{equation}
Following the existing literature 
(e.g., \citealp{mccandless2009bayesian,zigler2013model,zigler2014uncertainty}), we assume that 
the thresholds 
$q_1$ and $q_2$ are predetermined by the researcher prior to estimating the propensity score.

Alternatively, the propensity score may enter the outcome model as a continuous covariate,
\begin{equation}
g(Z_i, ps(Z_i,\gamma_i;\bm{\lambda})) = [Z_i', ps(Z_i,\gamma_i;\bm{\lambda})]',
\end{equation}
which can be viewed as a continuous analogue of propensity-score stratification 
\citep{rosenbaum1983central} . 
In the remainder of the paper, we focus on the subclassification specification~(\ref{strata}) to illustrate estimation and inference.

\section{Bayesian Estimation and Inference}

We adopt a Bayesian approach for the estimation and inference of treatment effects based on the 
proposed PS-LFM~(\ref{dr}). Compared with frequentist approaches, the Bayesian framework naturally 
incorporates uncertainty in both propensity score estimation 
(e.g., \citealp[]{mccandless2009bayesian,alvarez2021uncertain}) and factor selection 
(e.g., \citealp[]{samartsidis2020bayesian,pang2022bayesian}).

\subsection{Re-parameterization, shrinkage, and model selection}

Following the literature on Bayesian factor analysis (e.g., \citealp{geweke1996measuring}), 
we assume that the factor loadings are drawn from an i.i.d.\ multivariate normal distribution 
$\mathcal{N}(0, \Sigma_\gamma)$, where $\Sigma_\gamma = \mathrm{Diag}\{\omega_1^2,\ldots,\omega_r^2\}$ 
is an $(r\times r)$ diagonal matrix. The common factors are assumed to follow an i.i.d.\ 
multivariate normal distribution $\mathcal{N}(0,I_r)$, where $I_r$ denotes the identity matrix. 
The idiosyncratic error term $\varepsilon_{it}$ is assumed to be i.i.d.\ normal with mean zero 
and variance $\sigma^2$ 
\footnote{It is natural to extend the static factor specification to more flexible dynamic structures, 
such as random-walk factors or AR(1) processes \citep{pang2022bayesian}. More complex specifications 
for the error term, such as stochastic volatility (e.g., \citealp{belmonte2014hierarchical}), can also 
be incorporated. For notational simplicity, we consider a static factor structure with 
i.i.d.\ normal errors.} .

For model selection and estimation, we adopt a re-parameterization of the original 
Model~(\ref{dr}) following \citet{fruhwirth2010stochastic} and \citet{pang2022bayesian}. 
For notational convenience, define $\tilde{Z}_i = g(Z_i, ps(Z_i,\gamma_i;\bm{\lambda}))$, so 
that $\tilde{Z}_i$ represents an expanded vector of individual covariates augmented with the 
propensity score. We further decompose
$\beta_t = \beta + \xi_t$, where $\beta$ captures time-invariant effects and $\xi_t$ represents 
time-varying random deviations with mean zero. In the terminology of mixed-effects models, 
$\beta$ corresponds to fixed effects and $\xi_t$ to random effects. Although more flexible 
temporal dependence structures could be incorporated, 
as with the latent factors, we adopt the simpler hierarchical specification 
$\xi_t \overset{iid}{\sim} \mathcal{N}(0,\Sigma_\xi)$, where $\Sigma_\xi$ is diagonal, 
in order to highlight the role of propensity-score augmentation.

Because propensity-score subclassification may produce a high-dimensional expanded covariate 
vector $\tilde{Z}_i$, and because the number of latent factors is unknown, we re-parameterize 
both the random effects $\xi_t$ and the factor loadings $\gamma_i$ to facilitate shrinkage-based 
model selection. The resulting model can be written as:
\begin{equation}\label{modelallps}
\begin{gathered}
W_i = \mathbbm{1}\{Z_i'\lambda_z + (\omega_\gamma \cdot \tilde{\gamma}_i)'\lambda_\gamma + 
\nu_i \ge 0\}, \\
Y_{it}(0) =
g(Z_i,ps(Z_i,\omega_\gamma \cdot \tilde{\gamma}_i;\bm{\lambda}))'\beta +
g(Z_i,ps(Z_i,\omega_\gamma \cdot \tilde{\gamma}_i;\bm{\lambda}))'
(\omega_\xi \cdot \tilde{\xi}_t) +
(\omega_\gamma \cdot \tilde{\gamma}_i)' f_t + \varepsilon_{it}, \\
Y_{it} = \delta_{it}D_{it} + Y_{it}(0), \\
\tilde{\xi}_t \sim \mathcal{N}(0,I_p), \quad
\tilde{\gamma}_i \sim \mathcal{N}(0,I_r), \quad
f_t \sim \mathcal{N}(0,I_r), \\
\nu_i \sim \mathcal{N}(0,1), \quad
\varepsilon_{it} \sim \mathcal{N}(0,\sigma^2).
\end{gathered}
\end{equation}

To illustrate model selection, consider the factor component. For the $j$-th factor $f_{t,j}$, 
if the associated scaling parameter $\omega_{\gamma,j} = 0$, the corresponding factor is excluded 
from both the outcome and treatment assignment models, as 
$\omega_{\gamma,j} \cdot \tilde{\gamma}_{i,j} \equiv 0 \ \forall i$. Because the prior distribution 
for $\omega_\gamma$ includes zero in the interior of its support, this continuous shrinkage 
re-parameterization enables factor selection through posterior shrinkage.

It is possible that some latent factor loadings affect treatment assignment but not potential outcomes. 
In this case, such loadings do not constitute confounders and therefore do not need to be controlled 
for in the outcome model. However, if priors were placed directly on factor loadings 
$\gamma_i$ in the original joint model~(\ref{dr}), exact shrinkage to zero would generally not be 
achievable. The same reasoning applies to the random-effect component $\xi_t$.

In addition to shrinkage-based model selection, identification of the latent factor structure 
poses another challenge. This issue has been well documented in the interactive fixed-effects 
literature (e.g., \citealp{bai2009panel}). Let $\Gamma = (\gamma_1,\ldots,\gamma_N)'$ denote the 
$N\times r$ matrix of factor loadings and $F=(f_1,\ldots,f_T)'$ the $T\times r$ matrix of common 
factors. For any invertible $r\times r$ matrix $A$, we have 
$F\Gamma'=(FA^{-1})(A\Gamma')$, implying that only a rotationally equivalent representation of 
the factor structure is identified.

Frequentist approaches typically impose normalization conditions such as $F'F/T=I_r$ and 
$\Gamma'\Gamma$ being diagonal with distinct entries \footnote{In the Bayesian framework, because the 
prior distribution for the factors already assumes an identity covariance matrix $I_r$, an additional 
normalization proposed by \citet{geweke1996measuring} imposes an identification restriction on the 
factor loadings. Specifically, the loading matrix is partitioned as $\Gamma = (\Gamma_1, \Gamma_2)$, 
where $\Gamma_1$ is constrained to be lower triangular with positive diagnoal values.}. 
These normalization conditions are likewise 
assumed in \citet{forino2025propensity}, where treatment assignment is specified as a Logit model with 
factor loadings included as covariates.

However, recent Bayesian work on shrinkage-based factor selection 
(e.g., \citealp{bhattacharya2011sparse,samartsidis2020bayesian}) has primarily focused on imputing 
counterfactual outcomes 
and estimating treatment effects, treating latent factors largely as nuisance parameters without 
imposing explicit identification constraints. Moreover, updates of hyper-parameters in the shrinkage 
priors typically rely on draws of factor loadings from their conditional posterior distributions.

In our framework, however, the latent factors also enter the treatment assignment model, making their 
identification necessary for coherent estimation. In the next subsection, we propose an approximate 
Bayesian algorithm that addresses the trade-off between shrinkage for model selection and normalization 
of factor loadings for modeling the treatment assignment mechanism.

\subsection{Model feedback and approximate Bayesian inference}\label{mfp} 

The re-expressed PS-LFM~(\ref{modelallps}) implies that the joint likelihood of the treatment 
assignment and outcome model can be expressed as follows:
\begin{equation}\label{likelihood}
\begin{aligned}
&f(\bm{Y}(0)^{obs}, \bm{W} \mid \bm{Z}, \tilde{\Gamma}, F, \tilde{\bm{\Xi}}, \bm{\omega}, \bm{\lambda}, \beta) \\
=& \prod_{i=1}^{N} 
\Phi(Z_i' \lambda_z + \gamma_i' (\omega_\gamma\cdot\tilde{\gamma}_i))^{W_i}
\left[1-\Phi(Z_i' \lambda_z + \gamma_i' (\omega_\gamma\cdot\tilde{\gamma}_i))\right]^{1-W_i} \\
&\times \sigma^{-(NT-\sum_{i,t}D_{it})}
\prod_{D_{it}=0}
\exp\!\left\{-\frac{1}{2\sigma^2}
\sum_{D_{it}=0}
\left(y_{it}-\tilde{Z}_i'\beta
-\tilde{Z}_i'(\omega_\xi\cdot\tilde{\xi}_t)
-(\omega_\gamma\cdot\tilde{\gamma}_i)'f_t
\right)^2\right\},
\end{aligned}
\end{equation}
where $\tilde{\Gamma} = (\tilde{\gamma}_1, \ldots, \tilde{\gamma}_N)$ and $\tilde{\bm{\Xi}} = (\tilde{\xi}_1, \ldots, \tilde{\xi}_T)$ denote the matrices of factor loadings and 
time-varying coefficients, respectively, and $\bm{\omega} = (\omega_\gamma, \omega_\xi)$ denotes 
their associated scale parameters that impose shrinkage.

Assigning prior distributions to the parameters 
$\Theta=(\tilde{\Gamma}, F, \tilde{\bm{\Xi}}, \bm{\omega}, \bm{\lambda}, \beta)$ allows straightforward 
implementation of a fully Bayesian analysis by sampling from the joint posterior distribution
\begin{equation}
    f(\Theta \mid \bm{Y}(0)^{obs}, \bm{W}, \bm{Z}) \propto 
f(\bm{Y}(0)^{obs}, \bm{W} \mid \bm{Z}, \Theta) f(\Theta).
\end{equation} 
While conceptually appealing, this 
approach may adversely affect posterior inference due to what is commonly referred to as 
\emph{model feedback}. Specifically, when updating the normalized factor loadings 
$\tilde{\gamma}_i$, which appear in both the treatment assignment and outcome models, the posterior 
updates would condition jointly on observed control outcomes and treatment assignments. This issue 
has been discussed in \citet{zigler2013model}.

To illustrate this issue within the proposed model, we consider the posterior analysis of the 
factor loadings as an example.
The problem of model feedback arises because the influence of factor loadings on treatment adoption, 
captured by the parameters $\lambda_\gamma$, is not separately identified from the scale of the 
factor loadings ($\omega_\gamma$) themselves. For example, dividing $\lambda_\gamma$ by a scalar 
$\ell$ while multiplying the factor loadings by $\ell$ leaves the treatment assignment mechanism 
unchanged. However, such scaling ambiguity affects identification of the (normalized) factor loadings 
in the outcome model, potentially propagating non-identifiability to other parameters and leading to 
unstable estimation of counterfactual outcomes and, consequently, treatment effects. A similar issue 
arises for the propensity score, which is determined by the treatment assignment mechanism yet also 
enters the outcome model, creating analogous identification and inference challenges.

To mitigate this issue, rather than conducting a fully Bayesian analysis, we adopt an alternative 
procedure that cuts the feedback loop between the treatment assignment and outcome stages, yielding what 
is often referred to as an approximate Bayesian analysis. The key idea is that when parameters 
appear in both the treatment assignment and outcome models, one component of the likelihood is omitted 
when updating those parameters, producing approximate posterior draws.

For instance, the factor loadings $\tilde{\Gamma}$ are updated using only the outcome model, whereas the 
propensity score parameters are updated using only the treatment assignment model, even though the 
factor loadings enter the treatment assignment model and the propensity scores appear in the outcome 
model. This approximate Bayesian approach, together with the residual adjustment terms in the outcome 
model, has been shown to effectively address model feedback issues in Bayesian propensity score analysis 
\citep{zigler2013model,zigler2014uncertainty}, and is therefore particularly suitable for settings with 
unobserved confounding, such as the causal panel data models.

It is worth noting that, by omitting the treatment-assignment likelihood component when updating the 
factor loadings, the approximate Bayesian algorithm facilitates identification of $\omega_\gamma$ 
and, consequently, coherent modeling of the treatment assignment mechanism. Because both $\omega_\gamma$ 
and the latent factors $\tilde{F}$ are updated solely from the outcome model, we follow 
\citet{ando2022bayesian} and apply a ``post-processing'' rotation to the draws of $F$ 
and $\tilde{\Gamma}$\footnote{%
\citet{ando2022bayesian} propose an EM algorithm for estimating a panel Probit model with interactive 
fixed effects. Their procedure updates factors and loadings sequentially without imposing normalization 
restrictions during the EM iterations and applies a rotation after convergence to enforce standard 
identifying restrictions.}.

Let $\hat{\Gamma} = \tilde{\Gamma}\,\operatorname{diag}(\omega_\gamma)$, 
where $\operatorname{diag}(\omega_\gamma)$ is the $(r\times r)$ diagonal matrix whose diagonal 
entries are given by $\omega_\gamma$. Define
$
M = (\frac{1}{T}\tilde{F}'\tilde{F})^{1/2}
(\frac{1}{N}\hat{\Gamma}'\hat{\Gamma})
(\frac{1}{T}\tilde{F}'\tilde{F})^{1/2}
$.
Compute the eigen decomposition $M = QDQ'$, where $Q$ is orthogonal and $D$ is diagonal. Then define 
the rotated loadings and factors as 
$
\Gamma = \hat{\Gamma}(\frac{1}{T}\tilde{F}'\tilde{F})^{1/2}Q,
$ and 
$F = \tilde{F}(\frac{1}{T}\tilde{F}'\tilde{F})^{-1/2}Q$.
By construction, the rotated $(\Gamma, F)$ satisfy the normalization restrictions in 
\citet{bai2009panel}. We then insert the rotated $\Gamma$ into the treatment assignment model 
to update the parameters $\lambda_\gamma$ and $\lambda_z$.

The implementation procedure for the approximate Bayesian analysis is summarized in 
Algorithm~\ref{alg:cap}. For prior specification, we assign hierarchical shrinkage priors 
\citep{park2008bayesian} to $\beta$, $\omega_\gamma$, and $\omega_\xi$ to facilitate model selection, 
and conjugate normal priors to $\lambda_z$ and $\lambda_\gamma$. Details of the prior specifications 
and the Markov chain Monte Carlo (MCMC) algorithm are provided in~\ref{app2}.

\begin{algorithm}
  \caption{An Approximately Bayesian MCMC Sampler}\label{alg:cap}
  With the most updated parameters, at iteration ($h+1$): 
  \begin{enumerate}[noitemsep, topsep=0pt]
    \item Rotate $\tilde{F}$ and $\hat{\Gamma}$, as defined above, to obtain factor loadings $\Gamma$ that satisfy the normalization restrictions,
    
    \item Update coefficients $ \lambda_z $ and $\lambda_\gamma$ in the treatment assignment model and 
    update the propensity score for each unit,
    
    \item Update $\tilde{Z}_i$ given the updated propensity scores, 

    \item Jointly Update parameters $\beta$, $\omega_\xi$ and $\omega_\gamma$, 

    \item Update factor loadings $\tilde{\gamma}_i$ for $i \in \{1, \ldots, N \}$
    using only the outcome model, 
    
    \item Jointly Update time-varying random effects $\tilde{\xi}_t$ and latent factors 
    $ f_t $ for $t \in \{1, \ldots, T \}$,
    
    \item Update error variance $ \sigma^{2} $ in the outcome model,

    \item Update the hyperparameters in the corresponding shrinkage priors,
    
    \item Update counterfacutal outcome $ Y_{it}(0) $ 
    and individual treatment effect  $ \delta_{it} = Y_{it}- Y_{it}(0) $ 
    for observations under treatment ($D_{it} = 1$),

    \item Average over $\delta_{it}$ for observations under treatment as an estimate of the ATTs. 
   
    \end{enumerate}
    Repeat Steps above at the next iteration until the Markov chains converge.

  \end{algorithm}

\section{Monte Carlo Studies}\label{mcs}

We conduct a series of Monte Carlo simulations to investigate the properties of the proposed 
propensity score–augmented Bayesian factor model (PS-LFM) and to compare its performance with that 
of the canonical Bayesian factor model (DM-LFM), and some other model specifications, for treatment 
effect estimation. Our primary estimand is the average treatment effect on the treated,
and we evaluate estimator performance in terms of bias, root mean squared error (RMSE), the sampling 
standard deviation (Sampling SD) of the posterior mean as a point estimator, 
and the coverage rate of the 95\% posterior credible intervals.

In particular, we examine how the proposed method performs relative to alternative models under two 
scenarios: (i) when units are stratified into propensity score strata and the  
coefficients vary across strata, and (ii) when no such stratification is present, so that the 
DM-LFM is correctly specified and propensity score adjustment is unnecessary.

We consider a setting with $N = 200$ units observed over $T = 50$ periods with staggered treatment 
adoption. For the case involving propensity 
score stratification, simulated datasets are generated according to the following data-generating 
process (DGP). For each unit $i$, we generate two observed unit-level covariates, 
$Z_{i1}$ and $Z_{i2}$, which are drawn independently from standard normal distributions, 
and two latent factor loadings, $\gamma_{i1}$ and $\gamma_{i2}$, that satisfy the restrictions in 
for identification. Treatment assignment follows the probabilistic rule:
$$
W_i = \mathbbm{1}\{Z_{i1} + Z_{i2} + \gamma_{i1} + \gamma_{i2} + \nu_i \ge 0\},
$$
where $\nu_i \overset{i.i.d.}{\sim} N(0,1)$. Then corresponding propensity score is
$
\Pr(W_i = 1) = \Phi(Z_{i1} + Z_{i2} + \gamma_{i1} + \gamma_{i2})
$.

Among treated units, half are randomly designated as ``early adopters'', which receive treatment from 
period $t = 45$, while the remaining half are ``late adopters'', receiving treatment from $t = 48$. 
Units are stratified into three propensity score strata using thresholds of $0.3$ and $0.6$. 
Specifically, a unit is assigned to group 1 if its propensity score is less than $0.3$, to group 3 if 
its propensity score exceeds $0.6$, and to group 2 otherwise. Potential outcomes under control follow 
the latent factor specification:
$$
Y_{it}(0) = X_{it}'\beta_{(g)} + Z_{i(g)}'\xi_{(g)t} + \gamma_i' f_t + \varepsilon_{it},
$$
where $g \in \{1,2\}$ indexes the propensity score stratum. Without loss of generality, we set the 
group-specific time-varying coefficients to follow heterogeneous sinusoidal time trends.
The latent factors $f_t$ and idiosyncratic errors $\varepsilon_{it}$ are drawn independently from 
standard normal distributions. Treatment effects are set to zero for all treated units in 
post-treatment periods.

For model estimation, we allow up to five latent factors with shrinkage priors for model selection 
for both the DM-LFM and the proposed PS-LFM.
For comparison, we consider three alternative specifications. The first is an oracle model in which the 
true propensity scores, corresponding propensity score strata, and the true number of latent factors 
are assumed known, thereby eliminating uncertainty in propensity score estimation and model selection. 
We refer to this specification as ``Oracle''.
The second one is the doubly robust difference-in-differences estimator for staggered adoption settings 
proposed by \citet{callaway2021difference}, which we denote as ``CS-DID''. For the CS-DID estimator, we 
include only the time-invariant covariates in both the treatment assignment and outcome models.
The last estimator is the fixed effects counterfactual (``FEct'') estimator proposed by 
\citet{liu2024practical}, which imputes counterfactual outcomes using an interactive fixed effects 
model. For the FEct estimator, we set the number of factors equal to the true value. We conduct 500 
Monte Carlo replications, and the results are reported in Table~\ref{mc_res1}. In addition, 
a simulated example comparing the estimated propensity scores and 
dynamic treatment effects across the various model specifications described above is presented 
in~\ref{app3}. 

The Oracle estimator exhibits the lowest bias, close to zero, and its 95\% credible interval achieves 
coverage rates near the nominal level. This is expected, as the true propensity score, and hence the propensity score strata, is known by construction.
The proposed PS-LFM estimator has a larger bias than the Oracle estimator, which is reasonable because 
the propensity score must be estimated from a treatment assignment model in which the latent factor 
loadings, serving as unobserved confounders, are inferred from the outcome model. Nevertheless, the 
coverage rate of the 95\% credible interval remains close to the nominal level.

In contrast, both the DM-LFM and FEct estimators exhibit larger bias and RMSE relative to the PS-LFM, as 
they ignore the propensity score strata. Their coverage rates are also below the nominal level compared 
to the PS-LFM.
Finally, the CS-DID estimator performs worse than both the DM-LFM and FEct estimators. This is because 
the DGP incorporates a latent factor structure that favors latent factor–based estimators, whereas 
CS-DID is a doubly robust estimator that relies on observed covariates to model both the treatment 
assignment and the (difference in) outcome.


\begin{table}[h] 
  \centering
  \caption{Monte Carlo Study Results for DGP with Propensity Score Strata \label{mc_res1}}
  \begin{tabular}{l *{4}{S[table-format=1.3]}}
  \toprule  
  Model & {Bias} & {RMSE} & {Sampling SD} & {Coverage Rate} \\ 
  \hline
  Oracle     &0.006 &0.147 &0.147 &0.960 \\
  PS-LFM     &-0.036 &0.138 &0.134 &0.945 \\
  DM-LFM     &-0.100 &0.183 &0.153 &0.880 \\           
  CS-DID     &-0.907 &1.636 &1.365 &0.330 \\
  FEct       &-0.097 &0.298 &0.283 &0.915 \\
  \bottomrule  
 \end{tabular}
\end{table}

\section{Empirical Application} 

To illustrate the applicability of the proposed PS-LFM in social sciences study, we re-examine the 
effect of prior connections with Timothy Geithner on abnormal stock returns of financial firms during 
the first ten trading days following the announcement of his nomination as U.S. Treasury Secretary 
in November 2008. It is originally studied by \citet{acemoglu2016value}, as the value of political 
connections remains an important topic in the political economy literature.

In their analysis, the authors adopt two approaches to evaluate the effect of political 
connection on firm value. The key step is estimating abnormal returns, defined as the difference 
between observed daily returns and counterfactual returns that would have been realized had firms not 
possessed prior connections. First, they estimate counterfactual returns using a linear regression of 
firm returns on market returns, proxied by the S\&P 500 index. As a robustness check, they implement 
the synthetic control method (SCM; \citealp[]{abadie2010synthetic}), constructing counterfactual 
returns as convex combinations of returns from firms without political connections. The authors 
control for three firm-level covariates commonly used in asset pricing studies: firm size measured 
by total assets (on a log scale), profitability captured by return on equity (ROE), and leverage 
measured as the ratio of total debt to total capital in 2008 
\footnote{These firm-level covariate data are taken from  the Worldscope database.}. They also 
apply propensity score matching based on these covariates to improve covariate balance between 
treated and control firms.

Both approaches are grounded in factor-model perspectives on asset pricing 
\citep{geweke1996measuring}, suggesting that the proposed method, which incorporates latent factor 
structures, provides a flexible framework for constructing counterfactual returns. Moreover, 
concerns about unobserved firm-level confounders motivate our re-analysis. Financial firms with 
similar propensities for Geithner connection may share exposure to unobserved economic 
or political factors affecting stock returns, potentially differing systematically from firms with 
lower connection propensities. The proposed PS-LFM is therefore well suited to improve robustness 
in abnormal return estimation.

For the re-analysis, we consider a sample of 536 financial firms, among which 53 firms have prior 
connections with Timothy Geithner \footnote{Following \citet{acemoglu2016value}, a firm is considered 
connected if it satisfies at least one of three criteria: scheduled interactions with executives 
during 2007--2009, personal connections, or headquarters located in New York.}. The treatment 
indicator equals one for firms with such prior connections. The outcome variable is the daily stock 
return for each firm.

Consistent with the original study, we regard November 21, 2008, as the treatment date, with all 
treated firms receiving treatment simultaneously. The pre-treatment period includes all trading 
days in 2008 ending 30 days before the announcement, yielding 226 pre-treatment observations for 
model fitting. We then examine ten trading days following the announcement as the post-treatment 
period, resulting in eleven post-treatment observations in total.

For model specification, we control for the three firm-level covariates described above and allow 
their effects to vary over time and across propensity score strata. Because approximately 10\% of 
firms are treated, we use a threshold of 0.1 to form two propensity score strata. To account for 
unobserved factors in the return-generating process \citep{geweke1996measuring}, we include  
5 latent factors with shrinkage priors for model selection. 

\begin{figure}[!h]
\begin{center}
    \begin{tabular}{ccc}
         \includegraphics[scale = 0.25]{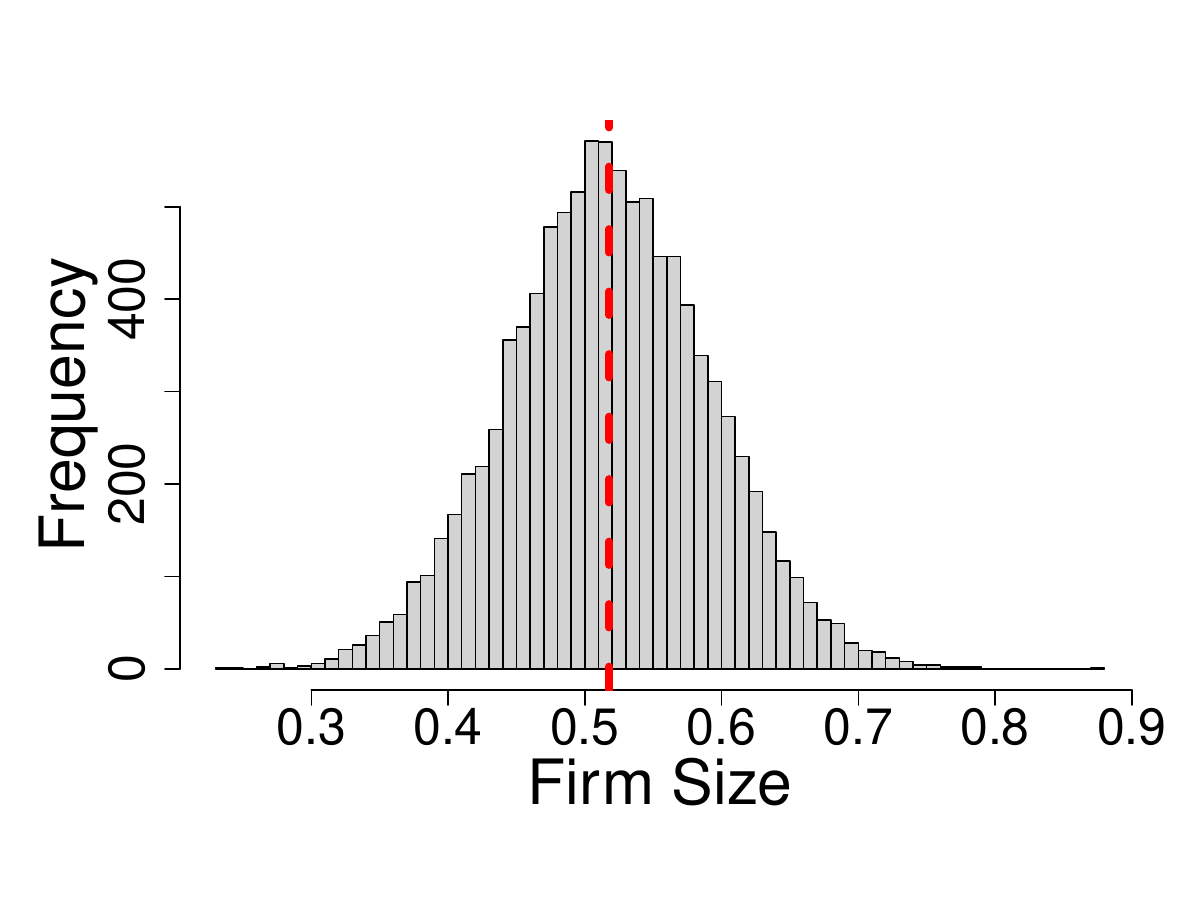} &
         \includegraphics[scale = 0.25]{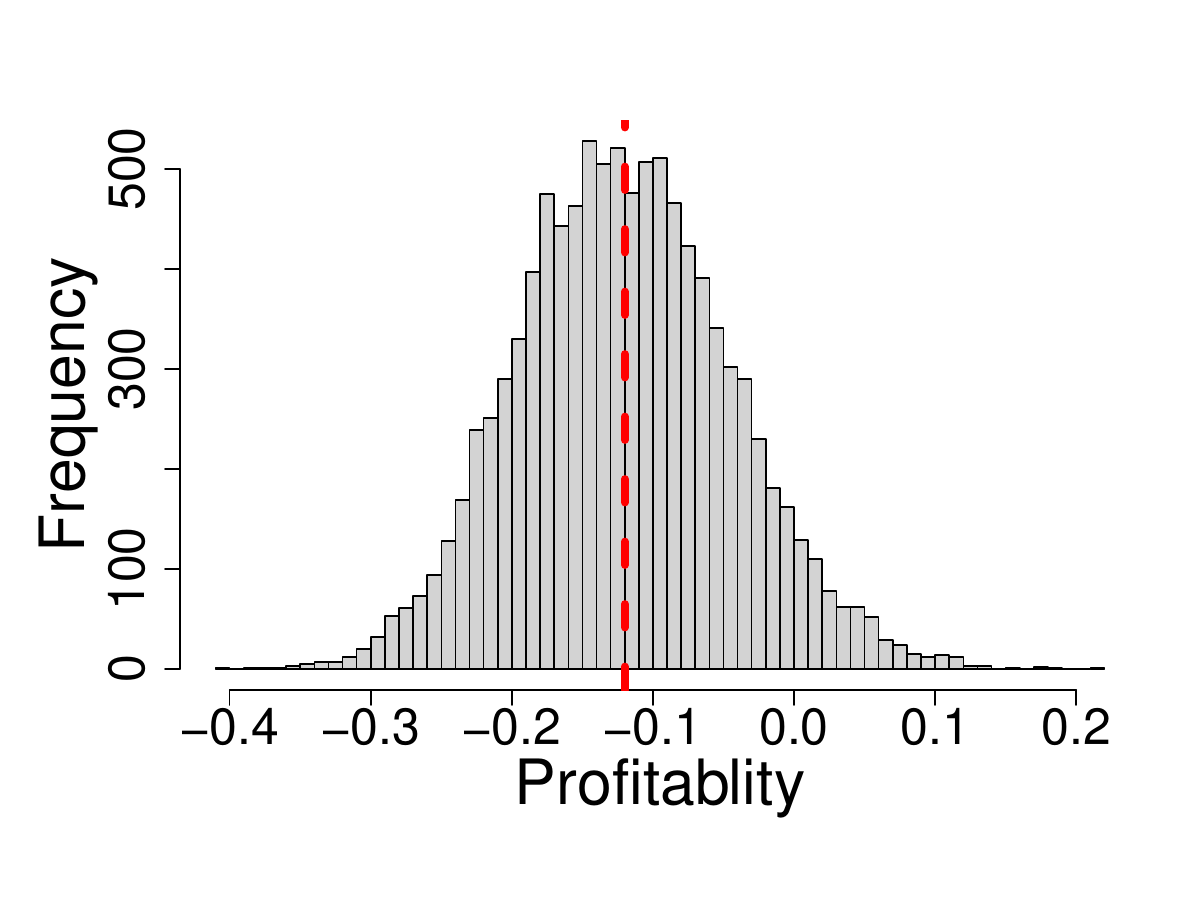} &
         \includegraphics[scale = 0.25]{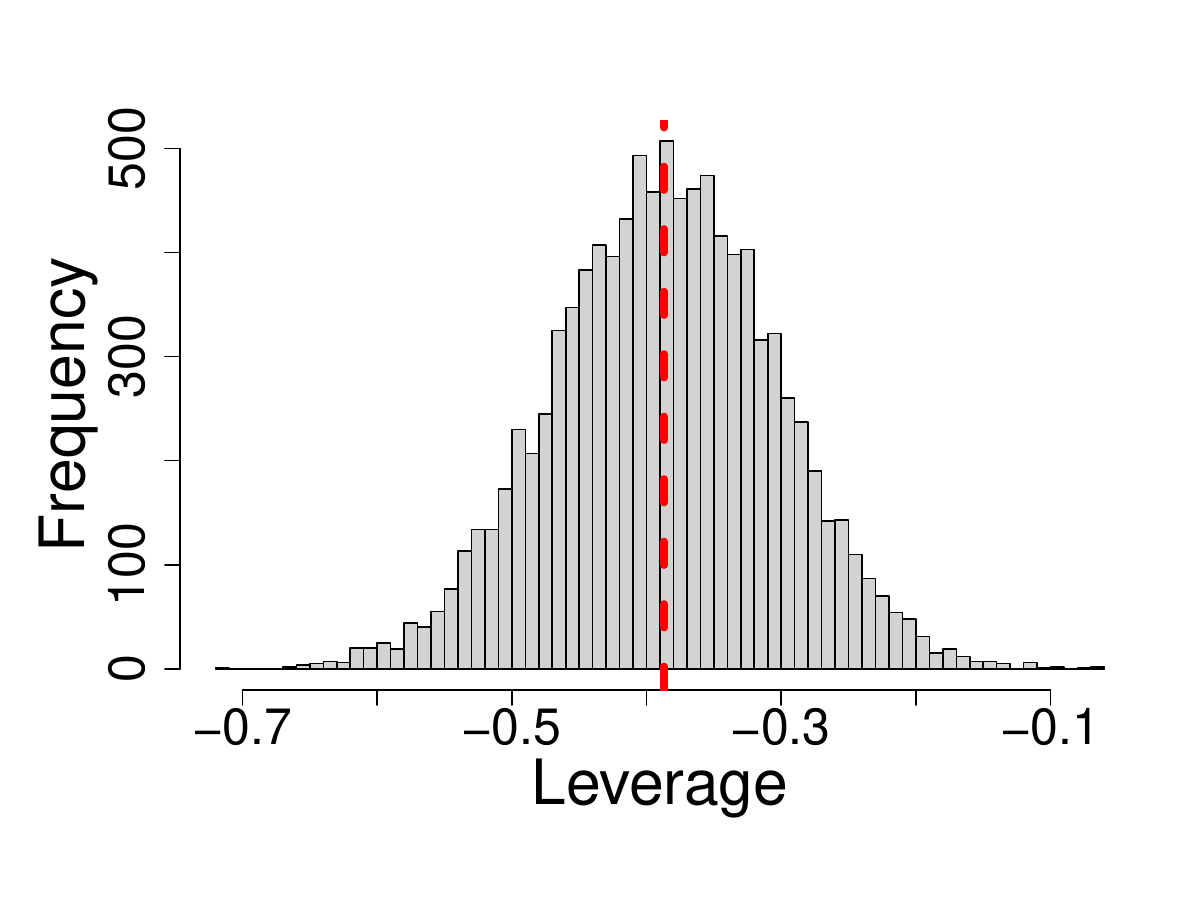} 
    \end{tabular}
    \caption{Firm level Covariates and Propensity of Geithner Connection}\label{ps_plot}
\end{center}
{\footnotesize {\it Note}: Histogram represents posterior distribution and red dashed line is the 
posterior mean.}
\end{figure}

The treatment assignment process is modeled such that the propensity for connection depends on both 
observed firm-level covariates and latent factor loadings. In addition, we control for the daily return 
of the S\&P 500 index as an observed common factor in the outcome model, allowing for firm-specific 
effects through a multi-level specification. The posterior distributions of the coefficients for the 
three firm-level covariates in the treatment assignment model, shown in Figure~\ref{ps_plot}, indicate 
that these covariates influence the propensity for connection with Timothy Geithner, consistent with 
the findings of the original study.

\begin{figure}[!h]
\begin{center}
    \begin{tabular}{c}
         \includegraphics[scale = 0.5]{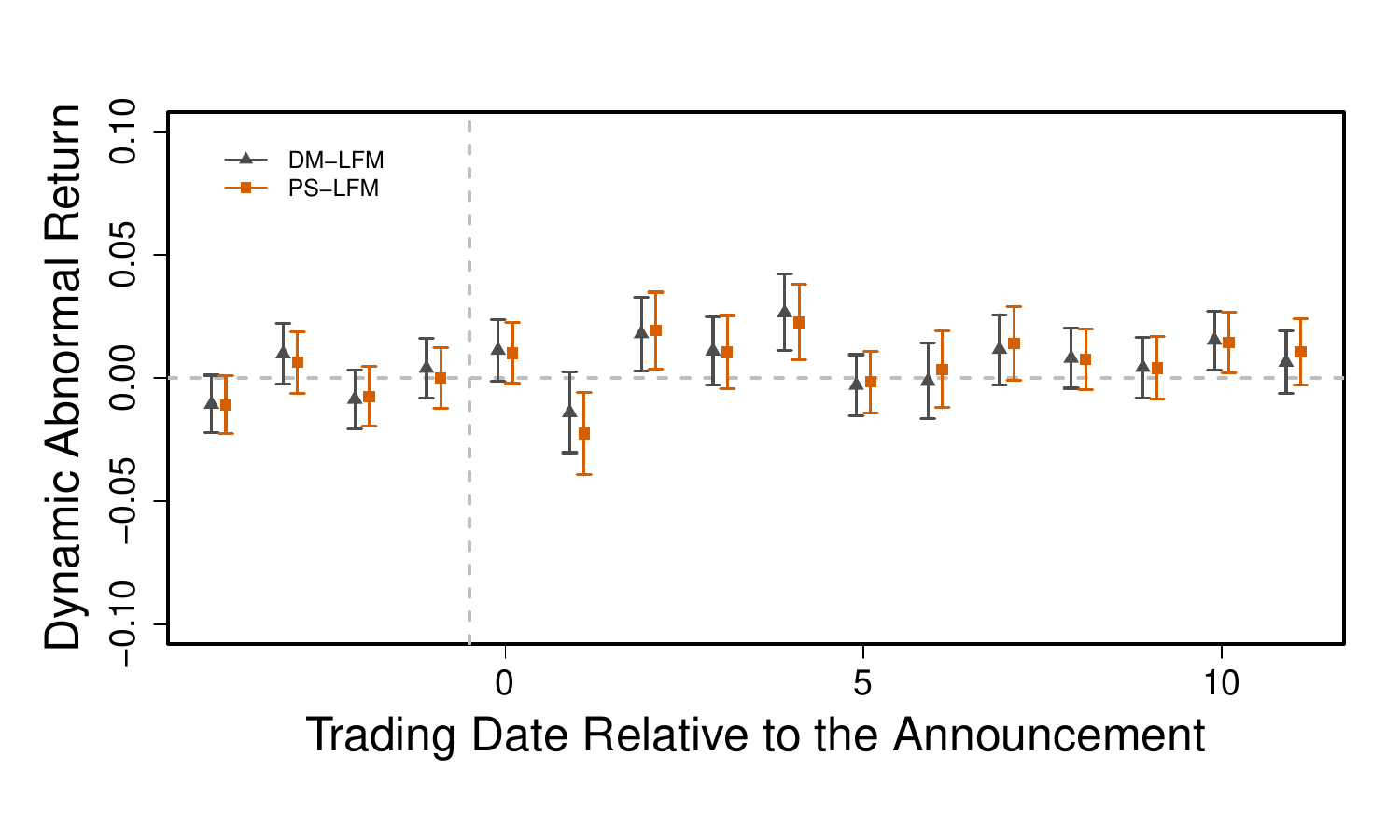}
    \end{tabular}
    \caption{Dynamic Effect of Geithner Connection on Abnormal Return}\label{datt}
\end{center}
{\footnotesize {\it Note}: The dots are the posterior means and the error bars represent corresponding  
95\% credible intervals.}
\end{figure}

To highlight the role of incorporating propensity scores together with latent factor loadings, 
we also estimate the DM-LFM as a benchmark for comparison. The estimated abnormal returns based on 
either model specification are displayed in Figure~\ref{datt}. Due to limited space, we only show 
four pre-treatment periods before the announcement. In general, the trajectories of abnormal returns 
based on either model specifications match closely to each other. However, for the announcement 
day, the 95\% CI of abnormal return based on the PS-LFM covers zero, while based on the DM-LFM does 
not cover zero. This also applies to the ninth trading day after the announcement, indicating slight 
yet non-negligible difference in abnormal return estimation when incorporating propensity score 
stratification.

As additional results, we examine the estimated time-varying effects of the three firm-level covariates. 
While firm size does not display significant heterogeneity across the two propensity score strata, 
profitability and leverage exhibit substantial heterogeneity in their time-varying effects. To further 
assess the validity of the estimates, we also conduct a placebo test by treating the two trading days 
prior to the announcement as placebo periods. The estimated placebo effects for both the proposed 
PS-LFM and the original DM-LFM are close to zero, and the corresponding 95\% CIs cover zero, which  
further supports the validity of the estimates. These additional results are reported in~\ref{EA}.

\section{Conclusion}

In this paper, we propose a novel Bayesian propensity score–augmented latent factor model for 
estimating treatment effects from time-series cross-sectional data. Compared with canonical 
causal panel data models that primarily emphasize outcome modeling, the proposed framework 
explicitly models the treatment assignment mechanism by incorporating latent factor loadings into 
the treatment assignment model. In addition, in the outcome stage, 
the potential outcome model incorporates the propensity score through stratification or other 
specifications like including it as an additional covariate. Relative to existing approaches, this 
framework provides greater flexibility and can capture additional heterogeneity across propensity score 
strata, which enables more credible comparisons between treated and control units within each stratum.

Although the Bayesian propensity score framework naturally incorporates uncertainty in propensity score 
estimation, it also introduces a model feedback problem because the propensity score appears in both the 
treatment assignment and outcome stages. To address this issue, we propose an approximate Bayesian 
algorithm in which factor loadings are estimated solely from the outcome model, after which appropriately 
rotated loadings are included as additional controls in the treatment assignment model. The propensity 
score is then estimated independently of the outcome model. This procedure mitigates the model feedback 
problem while balancing model selection considerations and identification of the latent factor structure.

Despite these advantages, the proposed method has several limitations. First, although the propensity 
score–augmented specification allows for heterogeneous effects across propensity score strata, it still 
relies on parametric assumptions for both the treatment assignment and outcome models. In particular, 
because the effect of the propensity score in the outcome model is assumed to follow a prespecified 
functional form, mis-specification, such as incorrect thresholds used for propensity score 
stratification, may lead to biased treatment effect estimates. Second, because latent factor loadings 
are estimated from the outcome model and subsequently incorporated into the treatment assignment model, 
the proposed approach does not fully satisfy the classical double-robustness property: correct 
specification of either the treatment or outcome model alone is insufficient for consistent estimation. 
Nevertheless, this dependence reflects the latent ignorability assumption required for identification 
in the presence of unobserved confounding. Future research aimed at 
addressing these limitations is worth pursuing.

\newpage
\bibliographystyle{chicago}
\bibliography{liter}

  \newpage
 \begin{center}
 {\large\bf SUPPLEMENTARY INFORMATION}
 \end{center}

 \appendix

 \section{MCMC Algorithm}\label{app2}
We provide the details of the proposed MCMC algorithm for sampling from the conditional 
posteriors of relevant parameters to estimate and make inference on the treatment effects. 
The priors for parameters to be estimated are as follows. We assign shrinkage priors for 
parameters $\beta$, $\omega_\xi$ and $\omega_\gamma$ in the outcome model, and weakly informative 
conjugate priors for $\lambda_z$ and $\lambda_\gamma$ in the treatment assignment model.
  \begin{itemize}

  \item $\lambda_z$ and $\lambda_\gamma$:
  \begin{equation}
  \begin{gathered}
  \lambda_z \sim \Nd(\bar{\lambda}_z, \B_z), \\
  \lambda_\gamma \sim \Nd(\bar{\lambda}_\gamma, \B_\gamma). \\
  \end{gathered}
  \end{equation}

  \item \( \beta \):
  \begin{equation}
  \begin{gathered}
  \beta_j | \tau_{\beta_j}^2 \sim \Nd(0, \tau_{\beta_j}^2), 
  \forall 1 \leq j \leq p \\
  \tau_{\beta_j}^2 | \kappa_{\beta} \sim Exp(\frac{\kappa_{\beta}^2}{2}) \\ 
  \kappa_{\beta}^2 \sim Gamma(a_1, a_2)
  \end{gathered}
  \end{equation}

  \item \( \omega_\xi \):
  \begin{equation}
  \begin{gathered}
  \omega_{\xi_j} | \tau_{\xi_j}^2 \sim N(0, \tau_{\xi_j}^2), 
  \forall 1 \leq j \leq p \\ 
  \tau_{\xi_j}^2 | \kappa_{\xi} \sim 
  Exp(\frac{\kappa_{\xi}^2}{2}) \\ 
  \kappa_{\xi}^2 \sim Gamma(c_1, c_2)
  \end{gathered}
  \end{equation}

  \item \( \omega_\gamma \):
  \begin{equation}
  \begin{gathered}
  \omega_{\gamma_j} | \tau_{\gamma_j}^2 \sim N(0, \omega_{\gamma_j}^2), 
  \forall 1 \leq j \leq r \\ 
  \omega_{\gamma_j}^2 | \kappa_{\gamma} \sim Exp(\frac{\kappa_{\gamma}^2}{2}) \\ 
  \kappa_{\gamma}^2 \sim Gamma(k_1, k_2)
  \end{gathered}
  \end{equation}

  \item \( \sigma^2 \):
  \begin{equation} 
  \sigma^{-2} \sim Gamma(e_1, e_2)
  \end{equation}
   \end{itemize}

 The steps of the Gibbs sampler are summarized as follows:
  
  \begin{enumerate} 
    

    \item Using the updated values of $\tilde{\gamma}_i$, $f_t$, and $\omega_\gamma$, 
    perform the rotation described in subsection~\ref{mfp} to obtain rotated $(\Gamma, F)$ that satisfy 
    the normalization restrictions in \citet{bai2009panel}. Then update 
    $\bm{\lambda} = (\lambda_z, \lambda_\gamma)$ in the treatment assignment model using the 
    latent outcome augmentation approach of \citet{albert1993bayesian}, and then the propensity score 
    $ps(Z_i,\omega_\gamma \cdot \tilde{\gamma}_i;\bm{\lambda})$ for each unit.

    \item Jointly update $\beta$, $\omega_\xi$ and $\omega_\gamma$ :
    
    Denote $$ 
    \tilde{Z}_{it} = 
    \left(g(Z_i,ps(Z_i,\omega_\gamma \cdot \tilde{\gamma}_i;\bm{\lambda})), \ 
    g(Z_i,ps(Z_i,\omega_\gamma \cdot \tilde{\gamma}_i;\bm{\lambda})) \cdot \tilde{\xi}_t, \ 
    \tilde{\gamma}_i \cdot f_t
    \right),
    $$
    \begin{equation}
    \begin{gathered}
    (\beta', \omega_\xi', \omega_\gamma')' \sim \Nd (\bar{\beta}, B_1), \\ 
    B_1 = \left(\sigma^{-2}\sum\limits_{D_{it} = 0} \tilde{Z}_{it} \tilde{Z}_{it}^{\prime} 
    + B_0^{-1} \right)^{-1}, \\
    \bar{\beta} = B_1\left(\sigma^{-2} \sum\limits_{D_{it} = 0} \tilde{Z}_{it} Y_{it} \right), \\
    B_0^{-1} = \text{Diag}( 
    \tau_{\beta_1}^{-2}, \ldots, \tau_{\beta_{p}}^{-2},
    \tau_{\omega_{\xi_1}}^{-2}, \ldots, \tau_{\omega_{\xi_p}}^{-2}
    \tau_{\omega_{\gamma_1}}^{-2}, \ldots, \tau_{\omega_{\gamma_r}}^{-2}).
    \end{gathered}
    \end{equation}

    \item Update \( \tilde{\gamma}_i \) for $i \in \{1, \ldots, N \}$ using only the outcome model
    \footnote{$I_r$ stands for ($r \times r$) identity matrix.}:

      Denote \( \tilde{f}_{t} = \tilde{\omega}_{\gamma} \cdot f_t \),
      \begin{equation}
      \begin{gathered}
        \tilde{\gamma}_i \sim \Nd (\bar{\gamma}, \Gamma_1), \\ 
        \Gamma_1 = \left(\sigma^{-2}
        \sum\limits_{t: D_{it} = 0} \tilde{f}_{t} \tilde{f}_{t}^{\prime} + 
        I_{r} \right)^{-1}, \\
        \bar{\gamma} = 
        \Gamma_1 \left(\sigma^{-2} 
        \sum\limits_{t: D_{it} = 0} \tilde{f}_{t} R_{it} \right), \\
        R_{it} = y_{it} - g(Z_i,ps(Z_i,\omega_\gamma \cdot \tilde{\gamma}_i;\bm{\lambda}))' \beta - 
        g(Z_i,ps(Z_i,\omega_\gamma \cdot \tilde{\gamma}_i;\bm{\lambda}))'
(\omega_\xi \cdot \tilde{\xi}_t).
      \end{gathered}
      \end{equation}

    \item Jointly update \( (\tilde{\xi}_t', f_t')' \)  for $t \in \{1, \ldots, T \}$:
    
      Denote $ \Psi_t =  (\tilde{\xi}_t', f_t')' $,
      $ \tilde{A}_{it} = \left(g(Z_i,ps(Z_i,\omega_\gamma \cdot \tilde{\gamma}_i;\bm{\lambda}))' \cdot \omega_{\xi}', \ 
      \omega_{\gamma}' \cdot \tilde{\gamma}_i'\right)' $, 
      \begin{equation}
      \begin{gathered}
        \Psi_t \sim N(\Omega_t^{-1}\mu_t, \Omega_t^{-1}) \\
        \mu_t = \sigma^{-2}\sum\limits_{t, D_{it} = 0} \tilde A_{it} U_{it},  \\
        \Omega_t = I_{(p + r)} + \sigma^{-2}\sum\limits_{t, D_{it} = 0} 
         \tilde A_{it} \tilde A_{it}^{\prime}, \\
        U_{it} = y_{it} -  X_{it}^{\prime}\beta. 
      \end{gathered}
      \end{equation}

    \item Update \( \tau_{\beta_j}^2 \)\footnote{Here ``IG'' stands for inverse Gaussian distribution.}: 
    \begin{equation}
    \tau_{\beta_j}^{-2} \sim IG(\sqrt{\frac{\kappa_{\beta}^2}{\beta_j^2}}, 
    \kappa_{\beta}^2), \quad \forall 1 \leq j \leq p
    \end{equation}

    \item Update \( \tau_{\xi_j}^2 \): 
    \begin{equation}
    \tau_{\xi_j}^{-2} \sim IG(\sqrt{\frac{\kappa_{\xi}^2}
    {\omega_{\xi_j}^2}, \kappa_{\xi}^2}), 
    \quad \forall 1 \leq j \leq p
    \end{equation}

    \item Update \( \tau_{\gamma_j}^2 \): 
    \begin{equation}
     \tau_{\gamma_j}^{-2} \sim IG(\sqrt{\frac{\kappa_{\gamma}^2}
     {\omega_{\gamma_j}^2}}, \kappa_{\gamma}^2), 
     \quad \forall 1 \leq j \leq r
    \end{equation}

    \item Update \( \kappa_\beta^2 \):
    \begin{equation}
    \kappa_\beta^{2} \sim Gamma(p + a_1, \frac{1}{2} 
    \sum\limits_{j=1}^{p} \tau_{\beta_j}^2 + a_2)
    \end{equation}

    \item Update \( \kappa_{\xi}^2 \):
    \begin{equation}
    \kappa_{\xi}^2 \sim Gamma(p + c_1, \frac{1}{2} 
    \sum\limits_{j=1}^{p} \tau_{\xi_j}^2 + c_2)
    \end{equation}

    \item Update \( \kappa_{\gamma}^2 \):
    \begin{equation}
    \kappa_{\gamma}^{2} \sim Gamma(r + k_1, \frac{1}{2} 
    \sum\limits_{j=1}^{r} \tau_{\gamma_j}^2 + k_2)
    \end{equation}

    \item Update \( \sigma^2 \):
    
    \begin{equation}
    \begin{gathered}
    \sigma^{-2} \sim Gamma(N_{obs} + e_1, \frac{1}{2} 
    \sum\limits_{D_{it}=0} (y_{it} - U_{it})^2 + e_2) \\
    N_{obs} = N \times T - \sum_{i=1}^{N}\sum_{t=1}^{T} D_{it}, \\ 
    U_{it} = g(Z_i,ps(Z_i,\omega_\gamma \cdot \tilde{\gamma}_i;\bm{\lambda}))'\beta +
g(Z_i,ps(Z_i,\omega_\gamma \cdot \tilde{\gamma}_i;\bm{\lambda}))'
(\omega_\xi \cdot \tilde{\xi}_t) +
(\omega_\gamma \cdot \tilde{\gamma}_i)' f_t.
    \end{gathered}
    \end{equation}

    \item Update the counterfacutal outcome \( Y_{it}(0) \) if \( D_{it} = 1 \):
    \begin{equation}
    \begin{gathered}
    Y_{it}(0) \sim N(R_{it}, \sigma^{2}), \\
    R_{it} =  g(Z_i,ps(Z_i,\omega_\gamma \cdot \tilde{\gamma}_i;\bm{\lambda}))'\beta +
g(Z_i,ps(Z_i,\omega_\gamma \cdot \tilde{\gamma}_i;\bm{\lambda}))'
(\omega_\xi \cdot \tilde{\xi}_t) +
(\omega_\gamma \cdot \tilde{\gamma}_i)' f_t.
    \end{gathered}
    \end{equation}

    \item Update \( \delta_{it} \) if \( D_{it} = 1 \):
    \begin{equation}
    \delta_{it} = Y_{it} - Y_{it}(0).
    \end{equation}
    
    \end{enumerate}
    
The proposed MCMC algorithm iterates through Steps 1–13 until convergence. Each of these steps 
corresponds to sampling from the conditional posterior distribution of the respective (cluster of) 
parameters given the remaining parameters and the observed data. Under standard regularity conditions, 
the resulting draws approximate samples from the joint posterior distribution after convergence.

In the final step, we impute the individual-level causal effects $\delta_{it}$ for treated observations. 
These draws are then summarized to obtain the average treatment effect on the treated  
\citep{brodersen2015inferring}.

\section{Additional Monte Carlo Studies}\label{app3}

\subsection{A Simulated Example}\label{app31}

\begin{figure}[!h]
\begin{center}
    \begin{tabular}{c}
         \includegraphics[scale = 0.45]{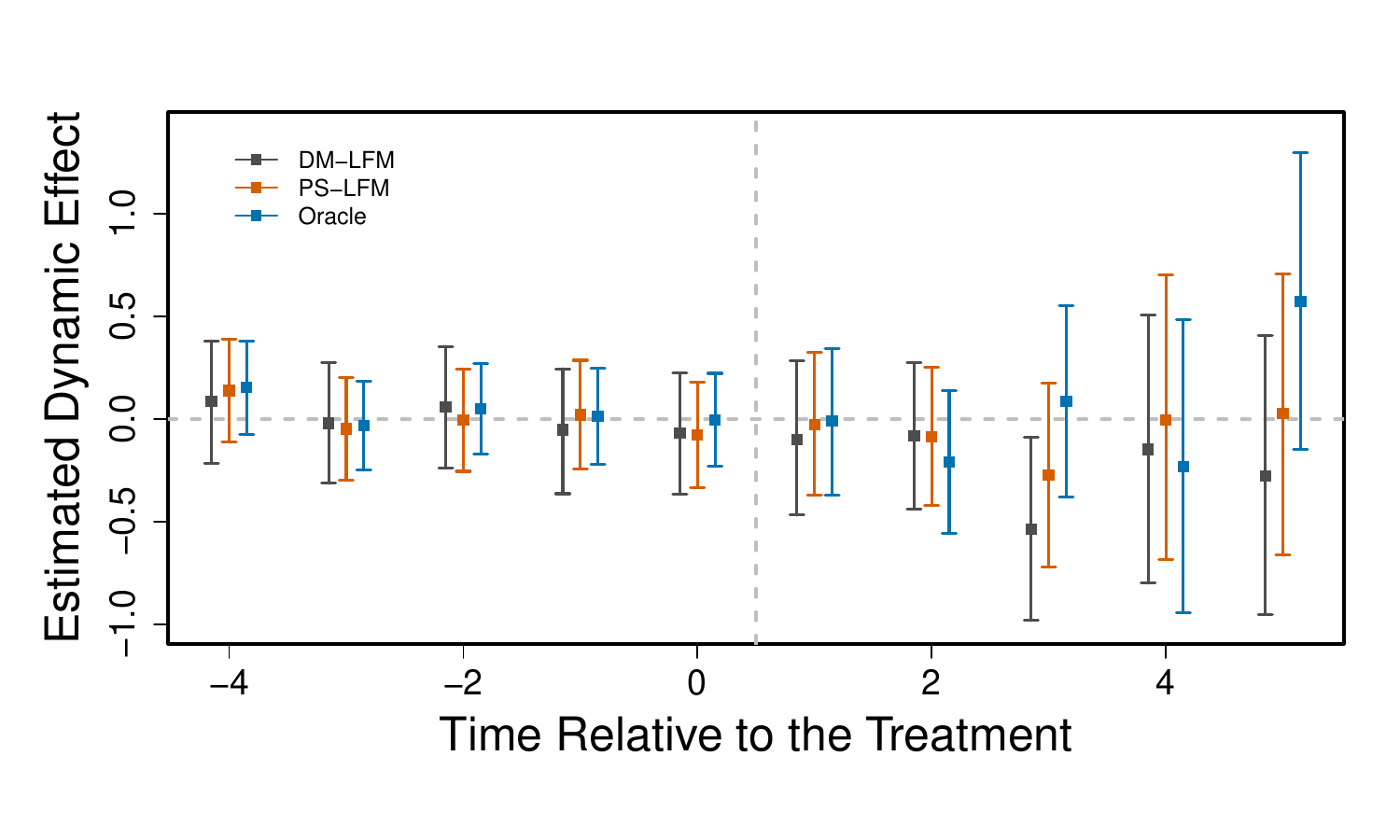} \\
         \includegraphics[scale = 0.45]{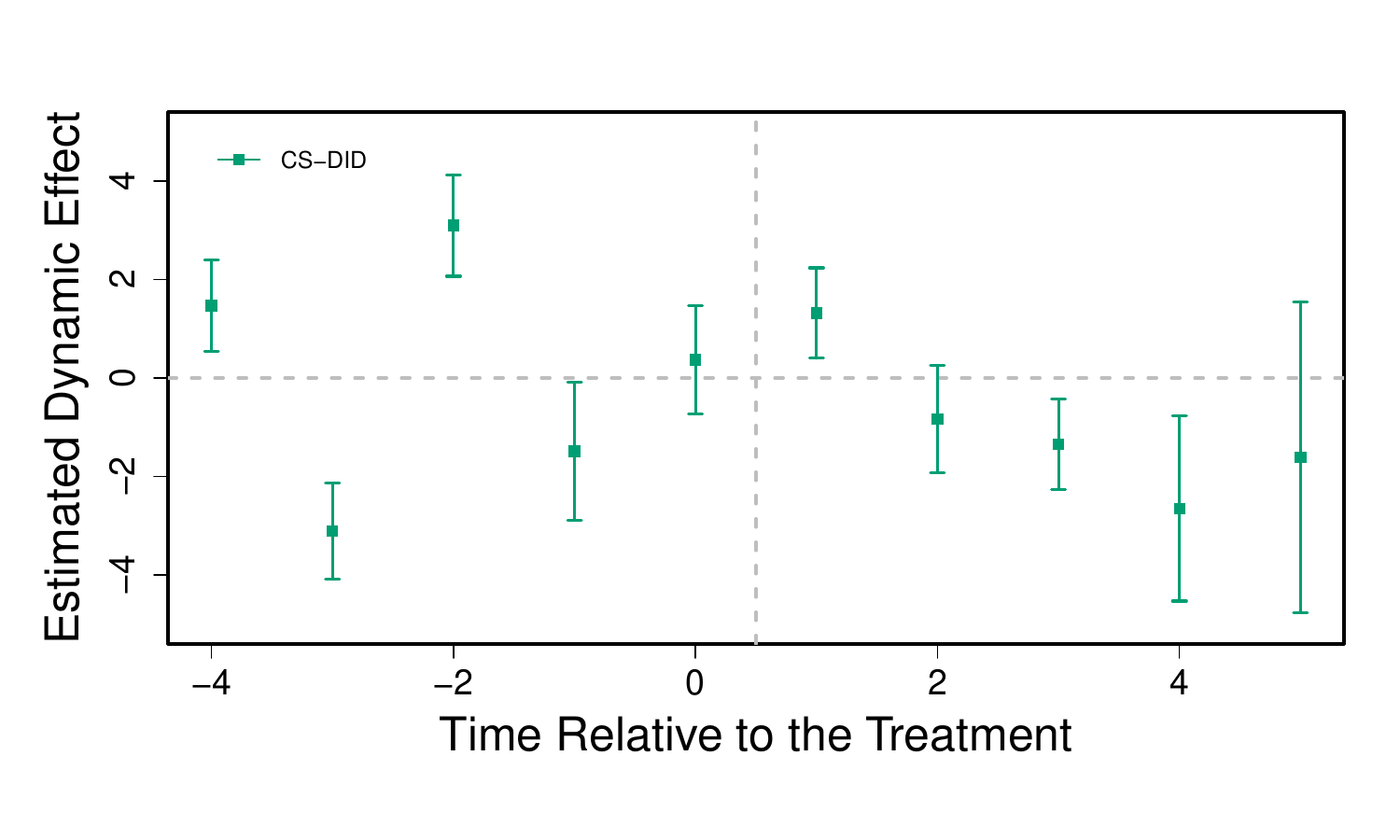} \\
         \includegraphics[scale = 0.45]{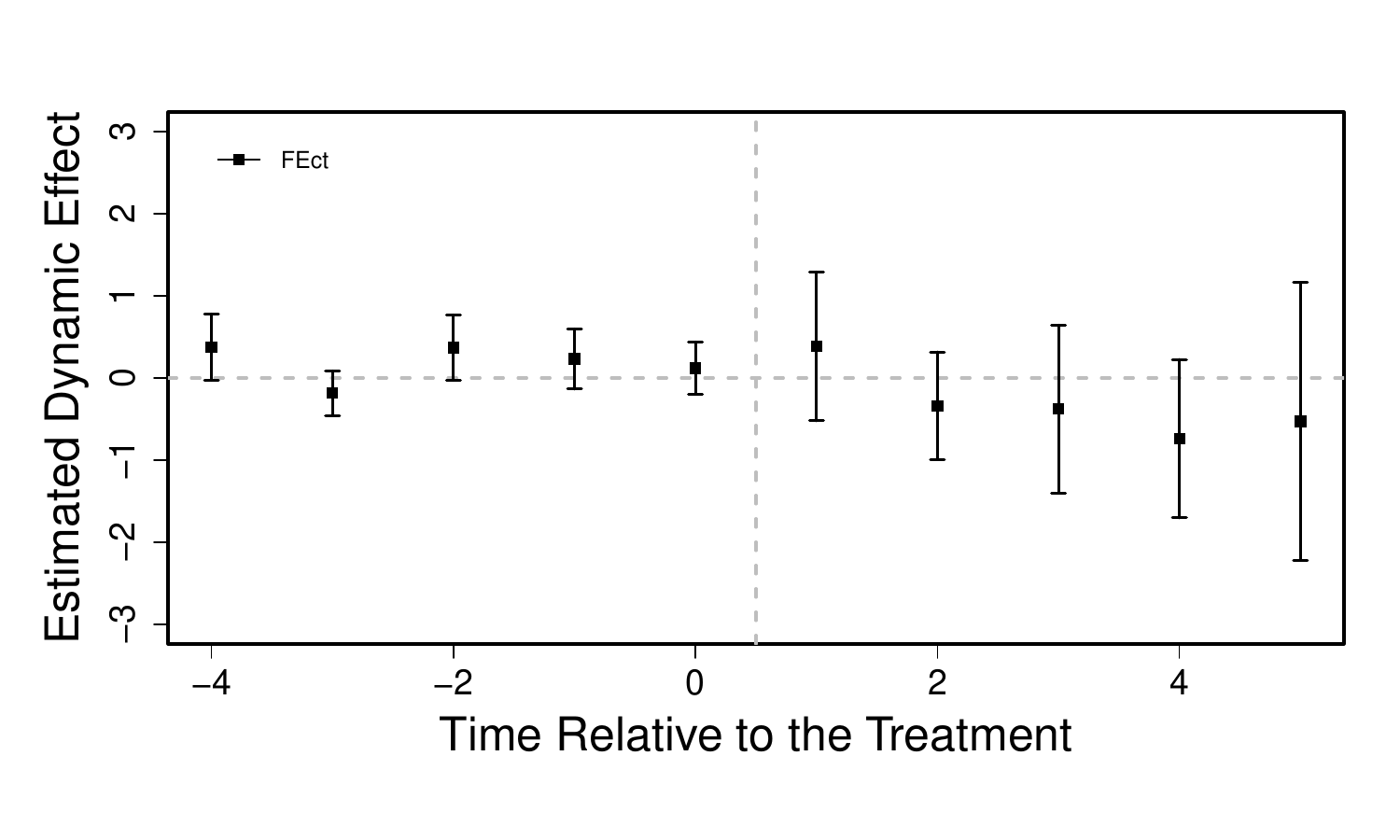} 
    \end{tabular}
    \caption{Estimated Dynamic Treatment Effects}\label{sim_plot1}
\end{center}
{\footnotesize {\it Note}: The dots are the posterior means and the error bars represent corresponding  
95\% credible intervals.}
\end{figure}

In this subsection, we apply the proposed method, along with alternative model specifications, 
to a simulated dataset generated from the same DGP with propensity score strata described in 
Section~\ref{mcs}, with $\delta_{it}=0$ for all $i$ and $t$. That is, $ATT_t \equiv 0$ for each 
post-treatment period. The estimated dynamic treatment effects for five pre-treatment periods 
and ten post-treatment periods are presented in Figure~\ref{sim_plot1}.

The upper panel displays the estimation results from the factor models. For the Oracle model, 
the posterior mean treatment effect is close to zero for both pre-treatment and post-treatment 
periods, and the 95\% credible intervals consistently cover zero. Similar patterns are observed 
for the proposed PS-LFM: posterior mean estimates remain close to zero across all periods, and 
the 95\% credible intervals cover zero throughout. However, these CIs are generally wider 
than those of the Oracle model, as additional uncertainty arises from propensity score 
estimation, propensity score stratification, and model selection.

By contrast, the DM-LFM exhibits evidence of bias. In particular, during the fourth 
post-treatment period, the 95\% credible interval does not include zero, suggesting bias induced 
by model mis-specification when propensity score stratification is ignored. Moreover, posterior 
mean estimates from the DM-LFM tend to deviate further from zero compared with both the Oracle 
model and the PS-LFM.

The lower panel reports results based on the CS-DID estimator. Several pre-treatment estimates 
are significantly different from zero, indicating potential violations of parallel trends due 
to unaccounted latent confounding. In addition, the confidence intervals are generally wider as we 
ignore the time-varying covariates in outcome model. 

\begin{figure}[!h]
\begin{center}
    \begin{tabular}{cc}
         \includegraphics[scale = 0.4]{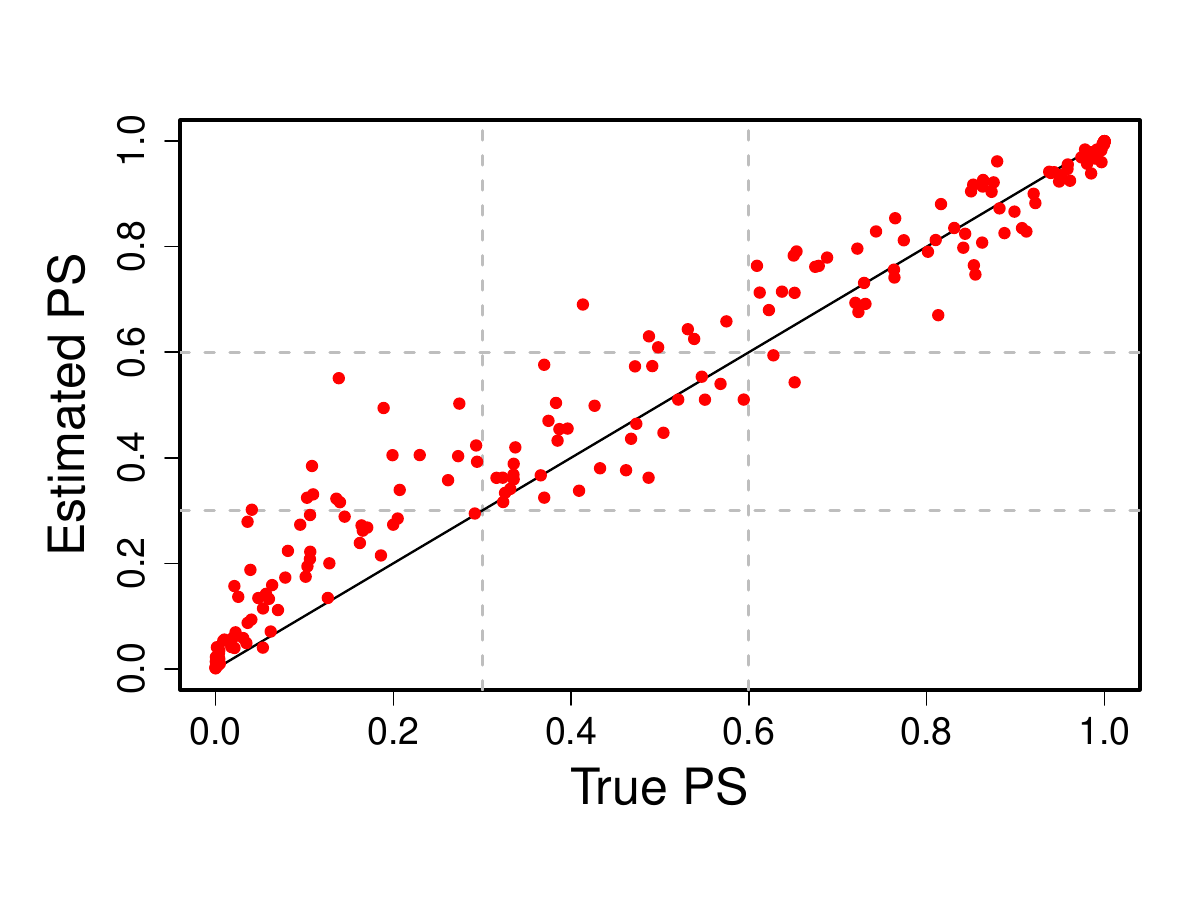} &
         \includegraphics[scale = 0.4]{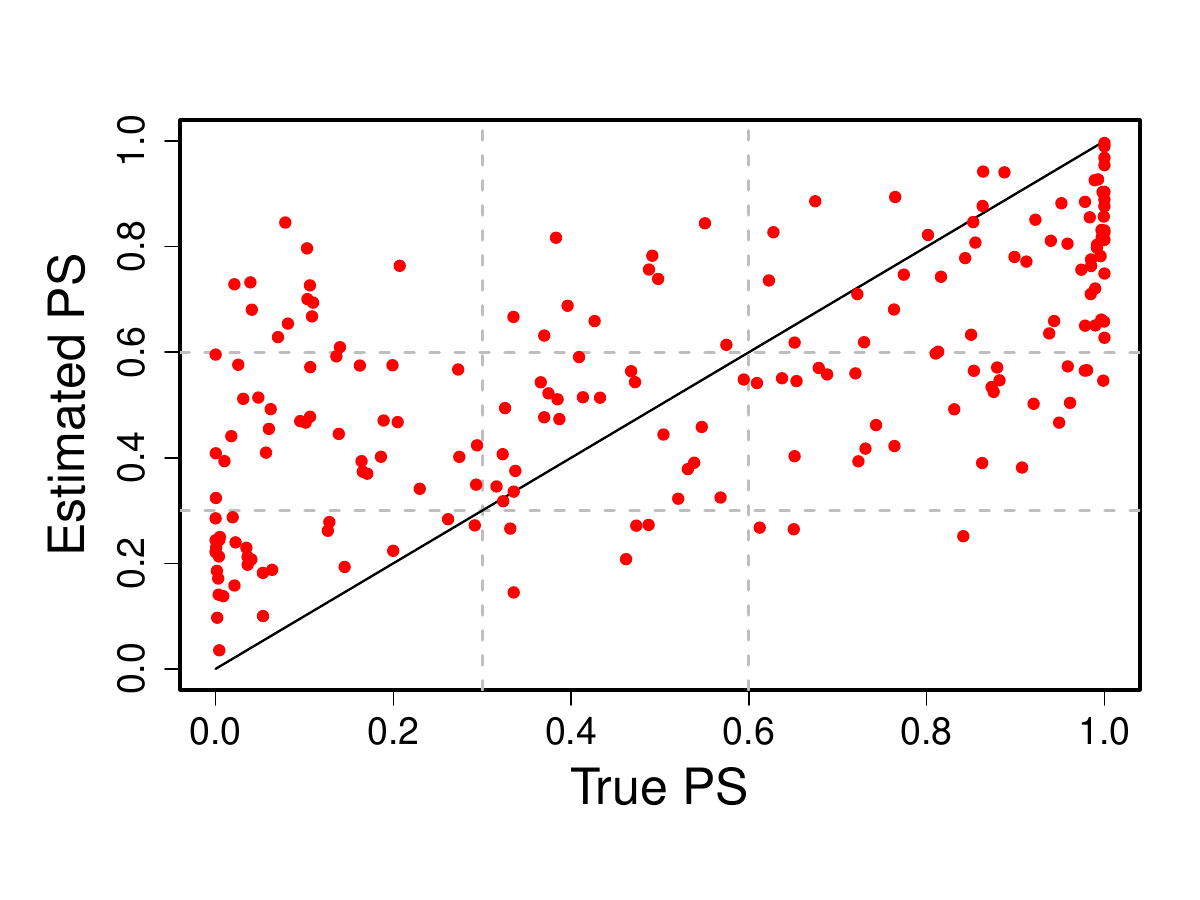} \\
         PS-LFM & Probit Model w/o Loadings \\
    \end{tabular}
    \caption{Estimated Propensity Scores with Factor Augmentation}\label{sim_plot2}
\end{center}
{\footnotesize {\it Note}: Posterior mean is used as the point estimate of propensity score. 
The black solid line represents the 45-degree line ($Y = X$). Grey dashed lines indicate 
the thresholds (0.3 and 0.6) that determine the propensity score strata.}
\end{figure}

We next assess the goodness of fit of the treatment assignment model under the PS-LFM, where 
factor loadings are estimated solely from the outcome model to mitigate the model feedback 
problem. Figure~\ref{sim_plot2} compares the estimated propensity scores with the true 
propensity scores. The left panel plots the estimated propensity scores from the PS-LFM against 
their true values. The estimates align closely with the 45-degree line ($Y=X$), indicating good 
fit of the treatment assignment model.

For comparison, the right panel shows propensity score estimates obtained from a Probit model 
that includes only observed unit-level covariates ($Z_{i1}$ and $Z_{i2}$) while ignoring latent 
factor loadings. In this case, the estimated propensity scores deviate substantially from the 
true values, suggesting that omitting latent confounders leads to poor model fit.

\section{Additional Results for the Empirical Application}\label{EA}

\begin{figure}[!h]
\begin{center}
    \begin{tabular}{cc}
         \includegraphics[scale = 0.4]{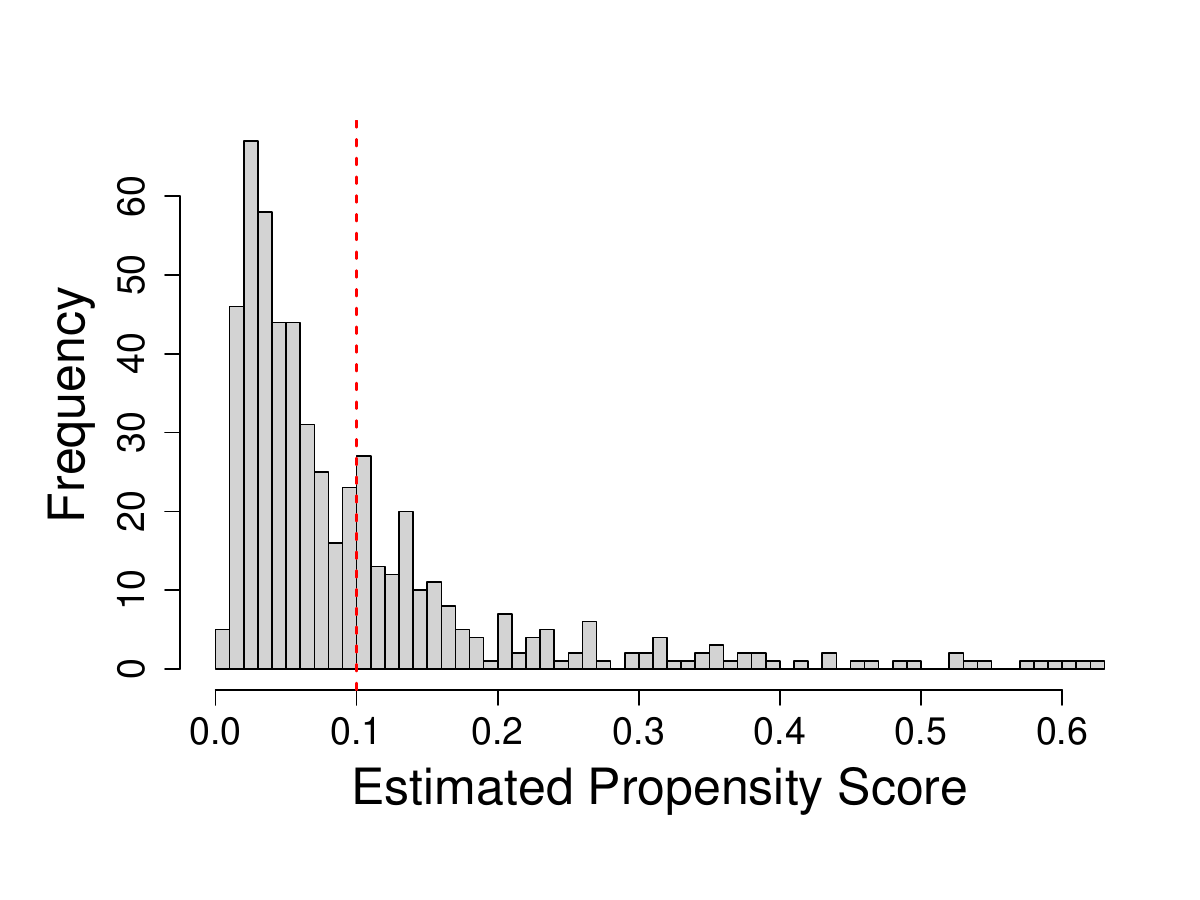} &
         \includegraphics[scale = 0.4]{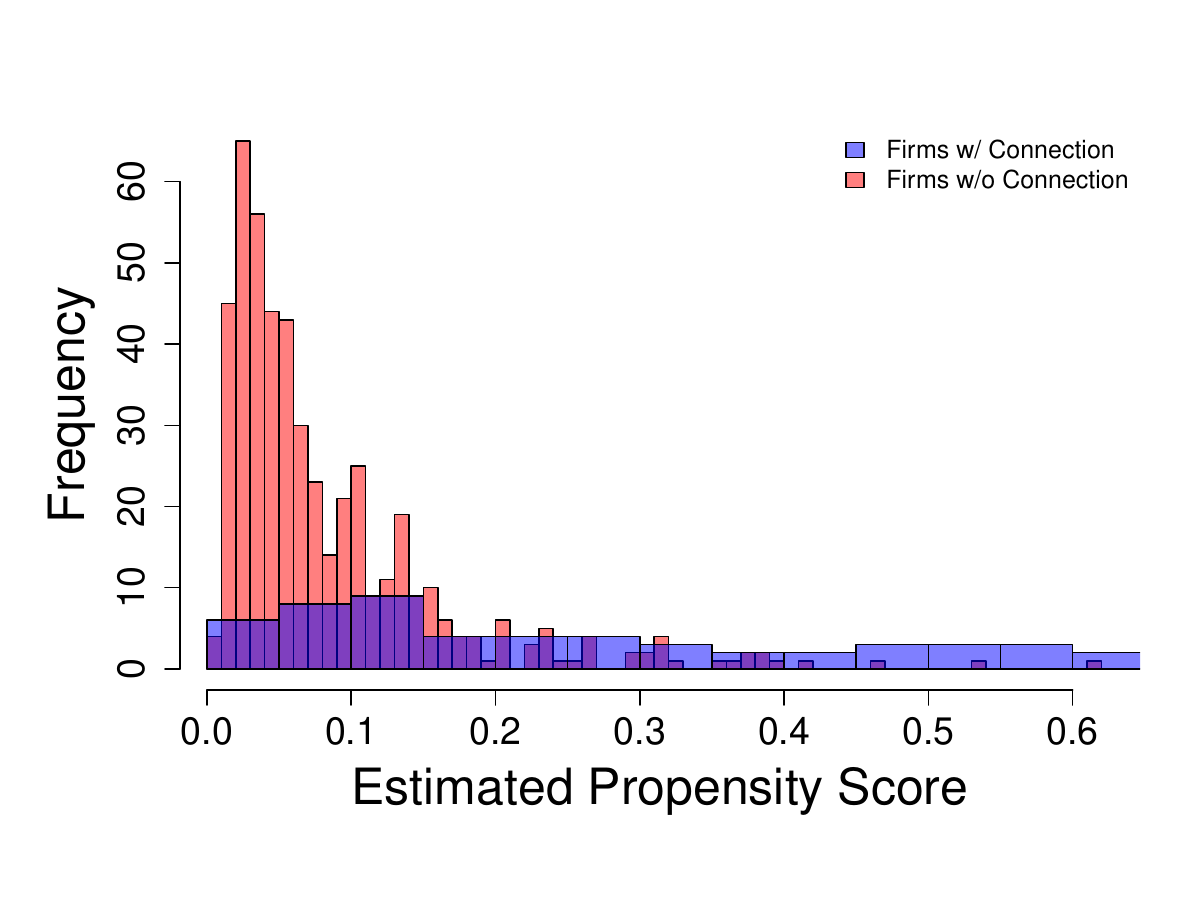} 
    \end{tabular}
    \caption{Distribution of Estimated Propensity Scores}\label{ps_dist}
\end{center}
{\footnotesize {\it Note}: Posterior mean is used as the point estimate of propensity score.}
\end{figure}

For additional results from the empirical application, we first examine the distribution of the 
estimated propensity scores, using the posterior mean as the point estimate. Figure~\ref{ps_dist} 
presents these distributions. The left panel shows that the estimated propensity scores are 
right-skewed, indicating that most financial firms have a relatively low propensity for connection 
with Timothy Geithner, while only a small fraction exhibit high connection propensity. The right 
panel compares the distributions separately for connected and unconnected firms. As shown, the 
propensity scores for connected firms are more dispersed than those for firms without connections.

\begin{figure}[!h]
\begin{center}
    \begin{tabular}{cc}
         \includegraphics[scale = 0.25]{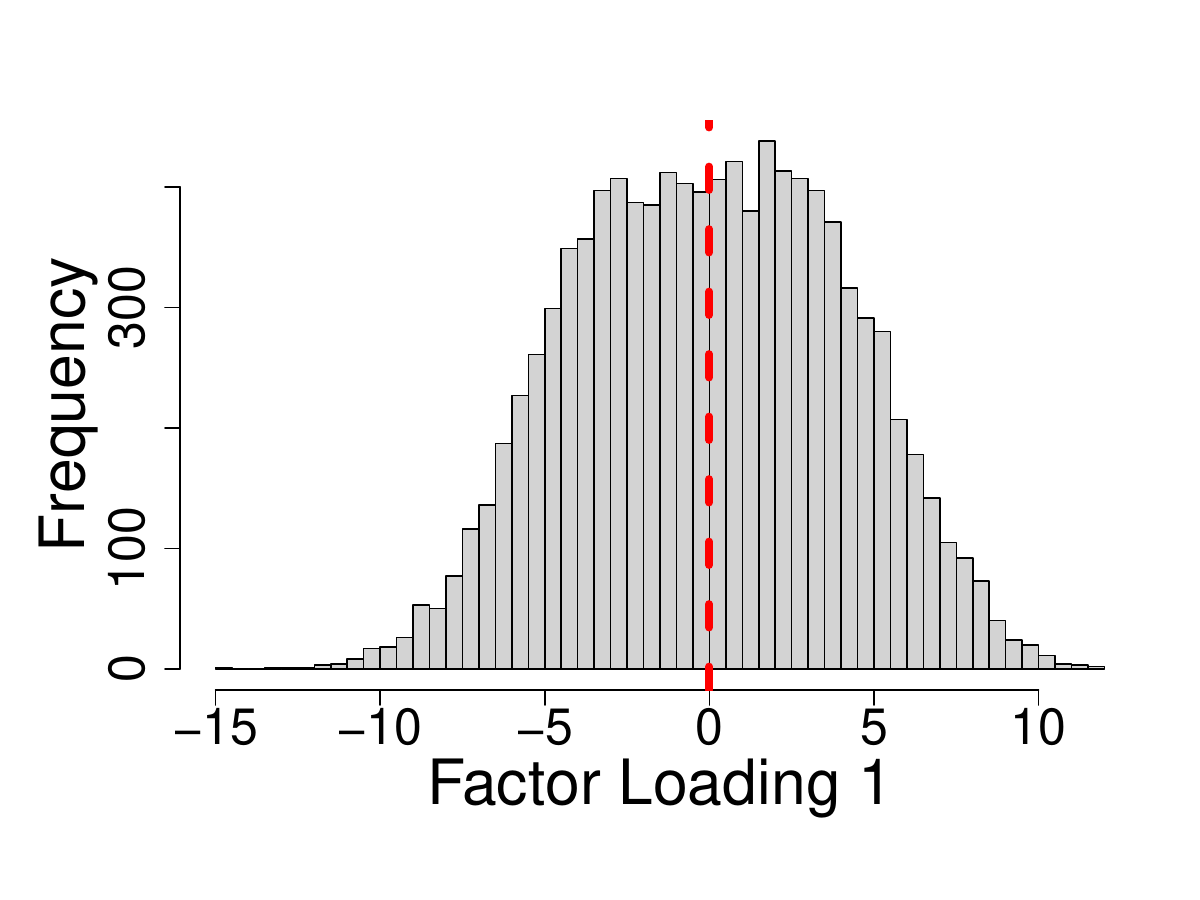} &
         \includegraphics[scale = 0.25]{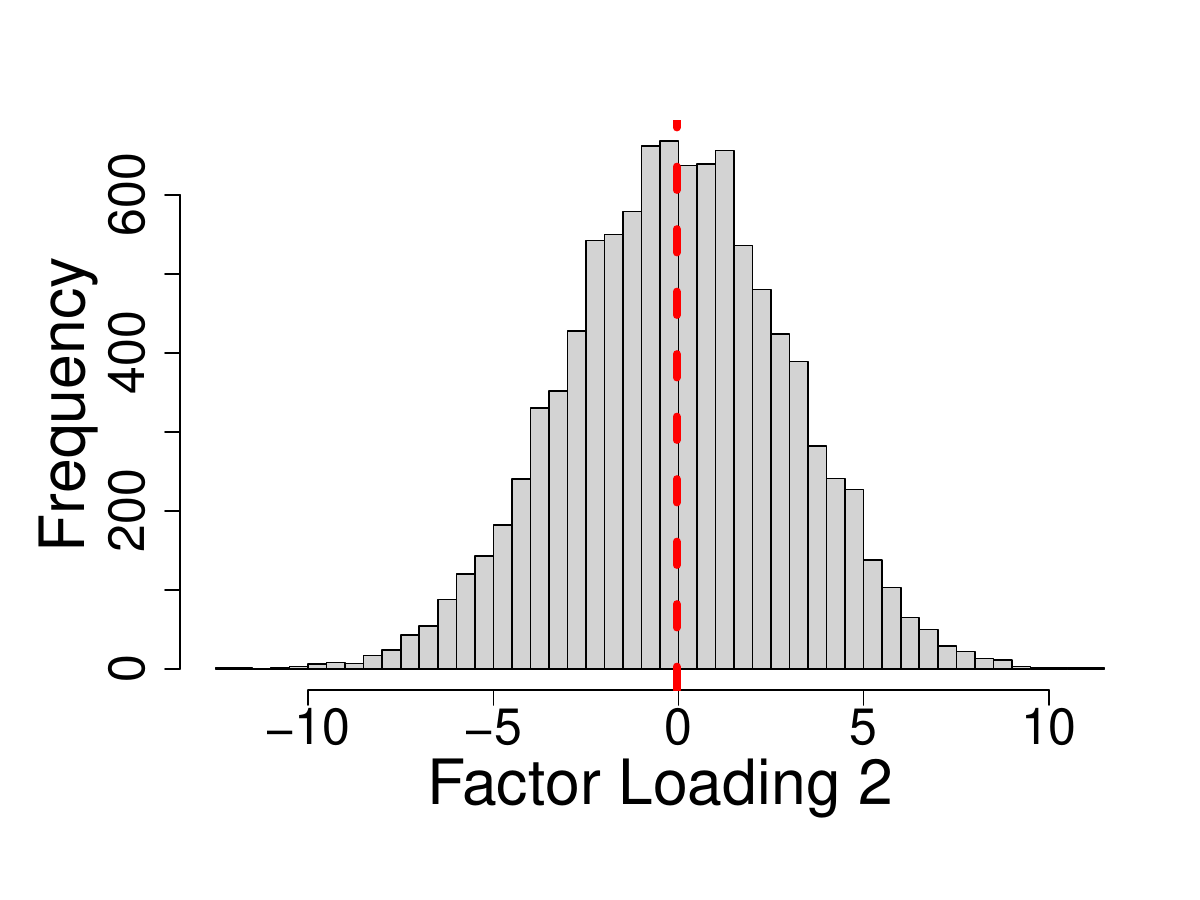} \\
         \includegraphics[scale = 0.25]{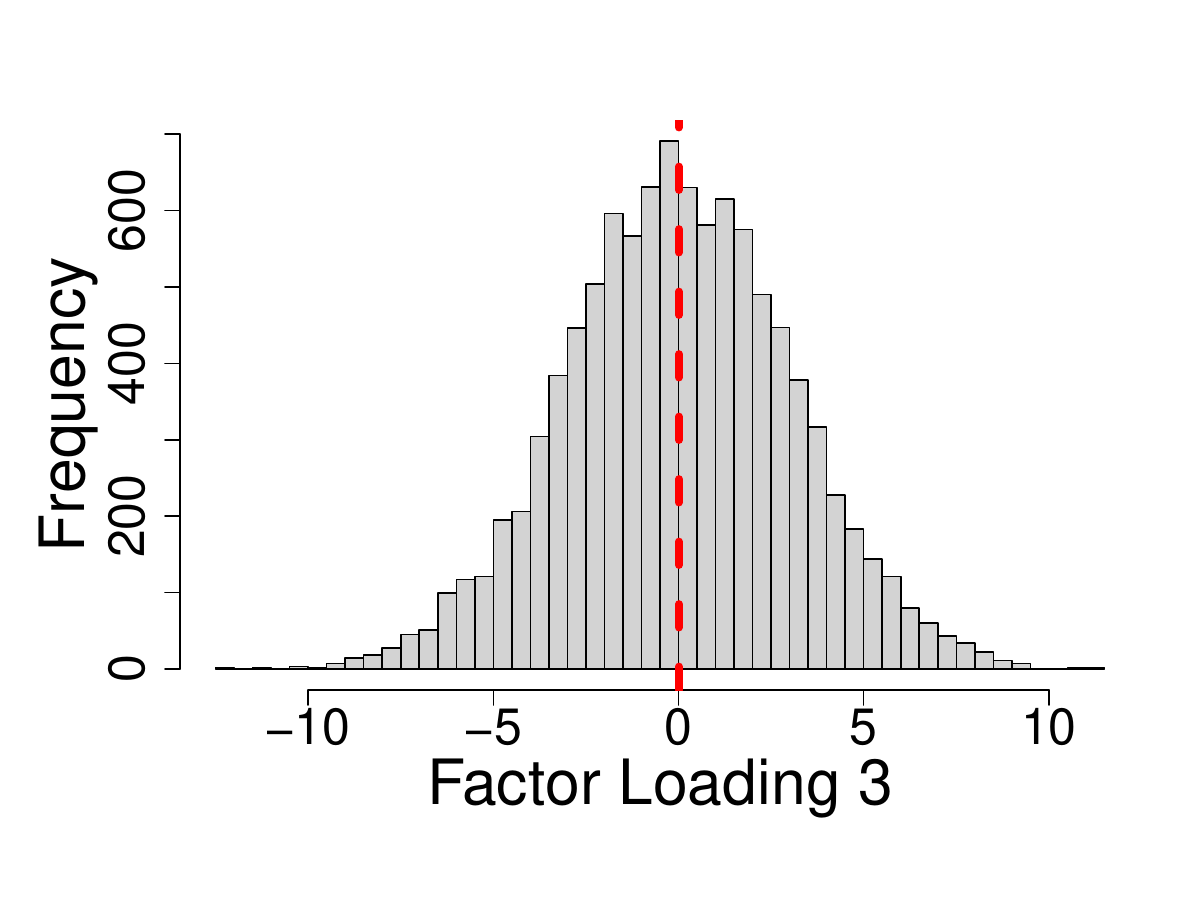} &
         \includegraphics[scale = 0.25]{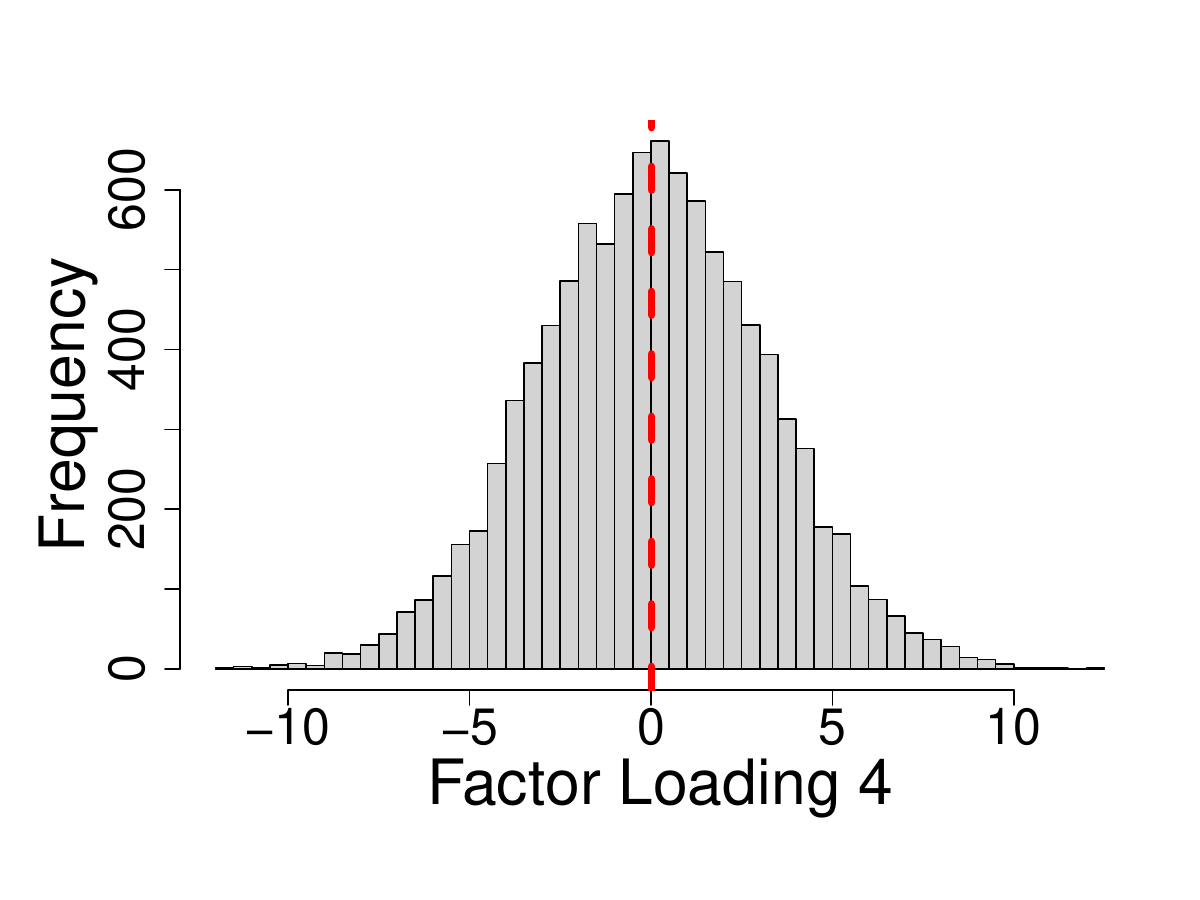} \\
         \includegraphics[scale = 0.25]{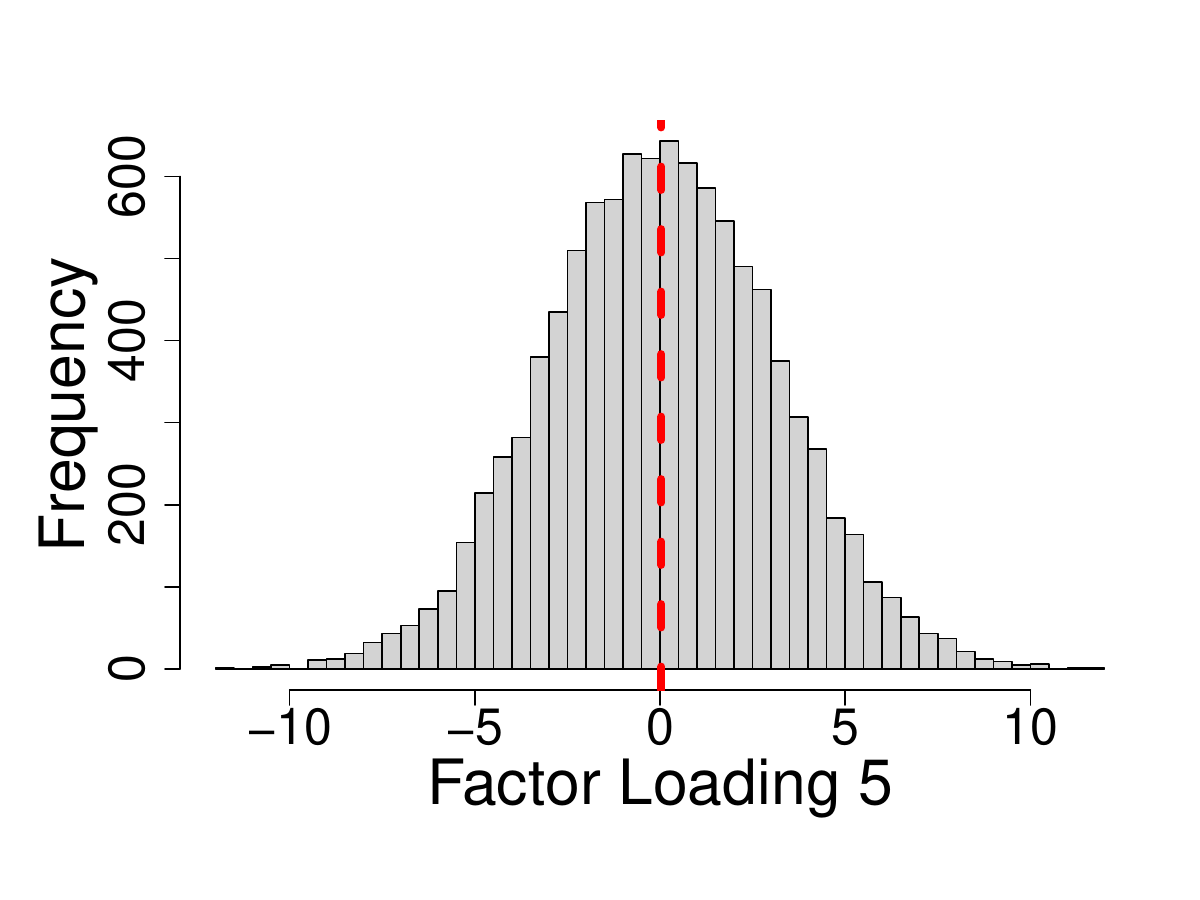} &
         \includegraphics[scale = 0.25]{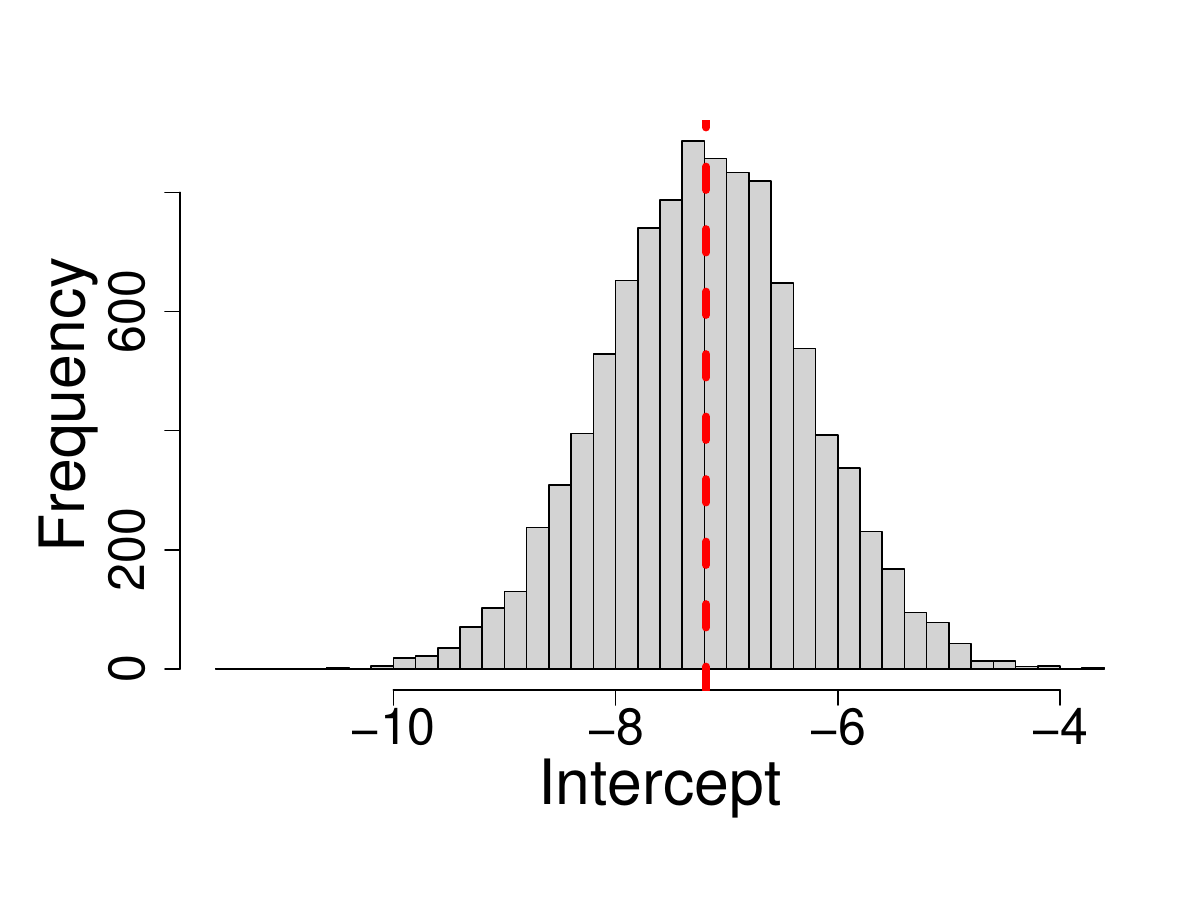}  \\
    \end{tabular}
    \caption{Factor Loadings and Propensity of Geithner Connection}\label{ps_plot2}
\end{center}
{\footnotesize {\it Note}: Histogram represents posterior distribution and red dashed line is the 
posterior mean.}
\end{figure}

Figure~\ref{ps_plot2} displays the posterior distributions of the coefficients associated with each 
factor loading, along with the intercept, in the treatment assignment model. The posterior means of 
the factor-loading coefficients are close to zero, suggesting limited additional latent confounding 
after controlling for firm size, profitability, and leverage.

\begin{figure}[!h]
\begin{center}
    \begin{tabular}{c}
         \includegraphics[scale = 0.45]{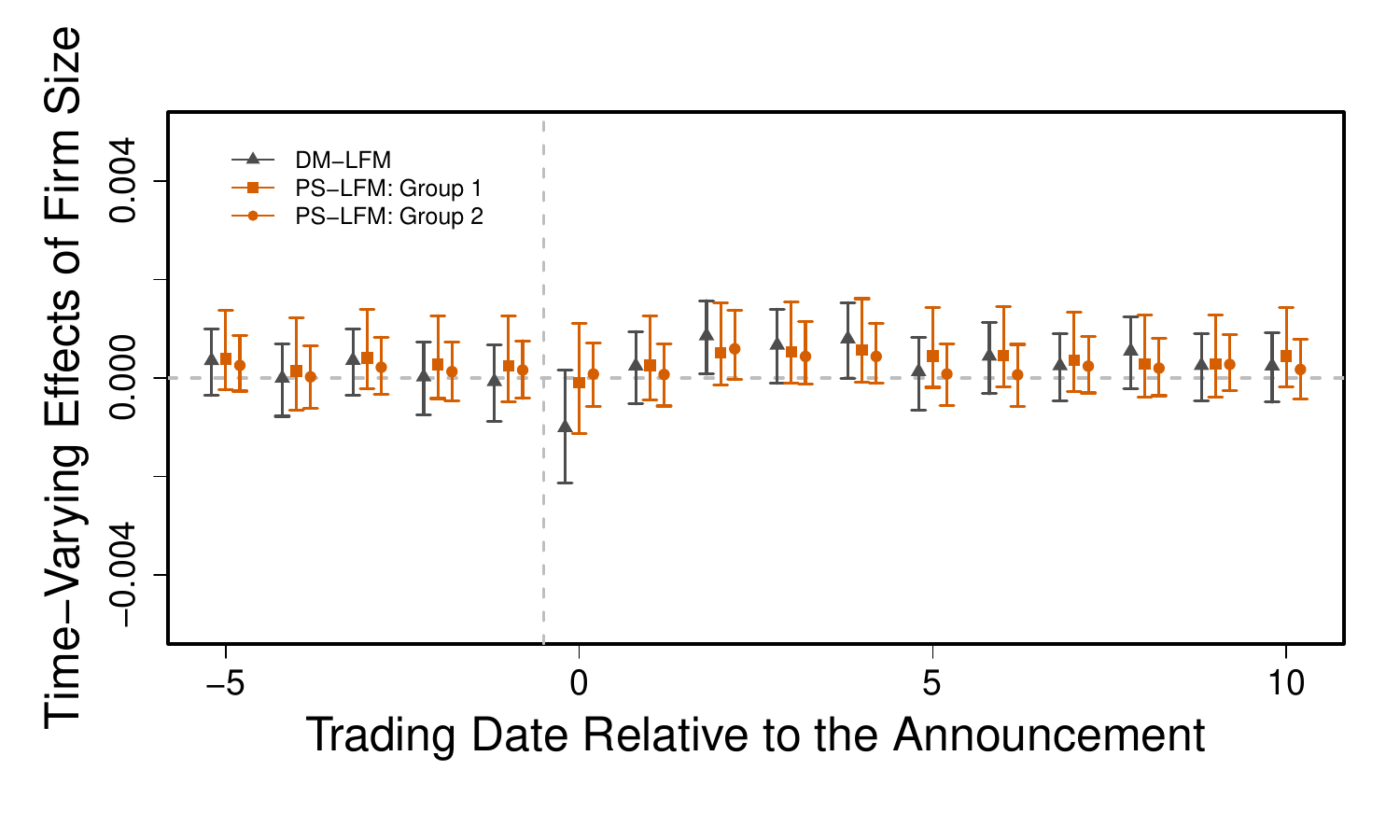} \\
         \includegraphics[scale = 0.45]{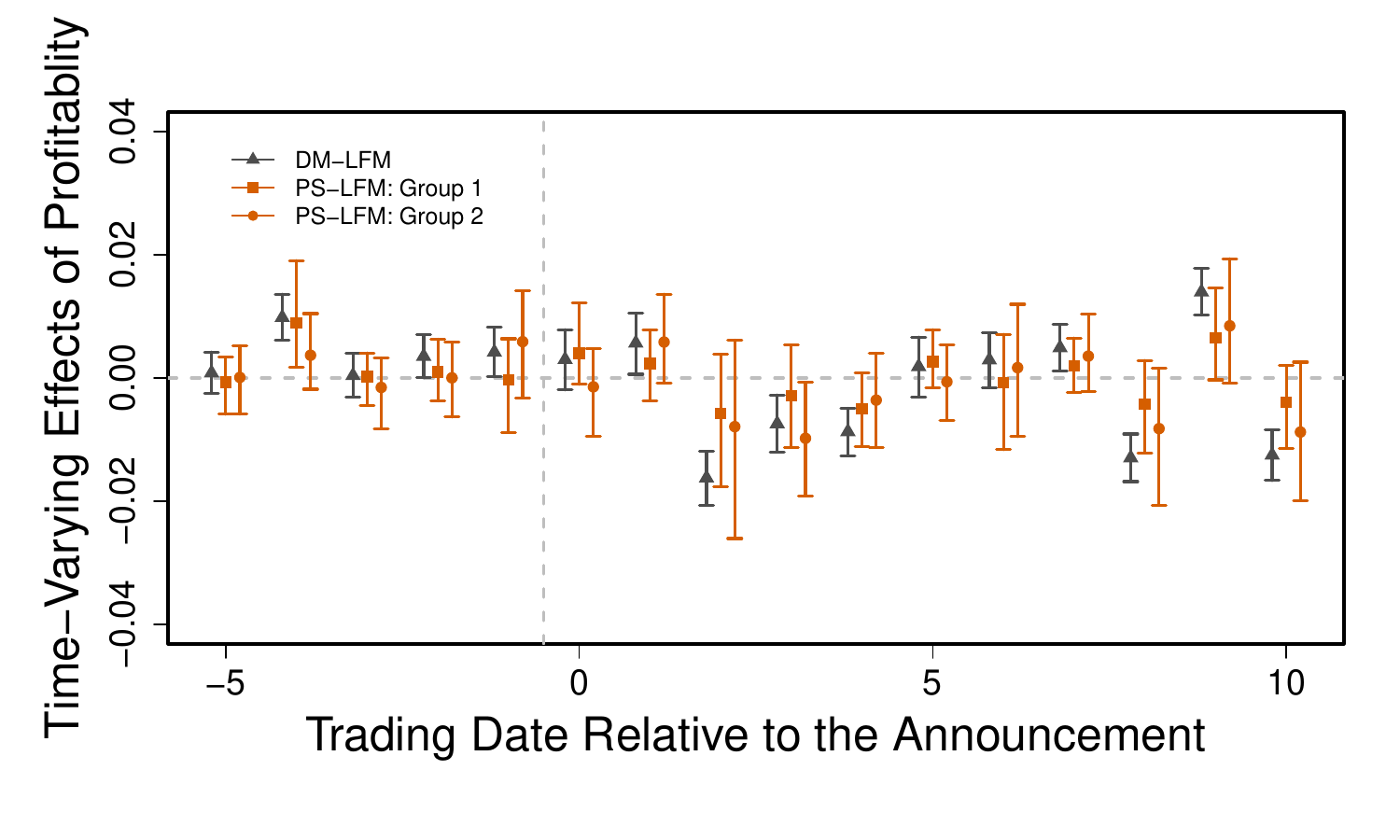} \\
         \includegraphics[scale = 0.45]{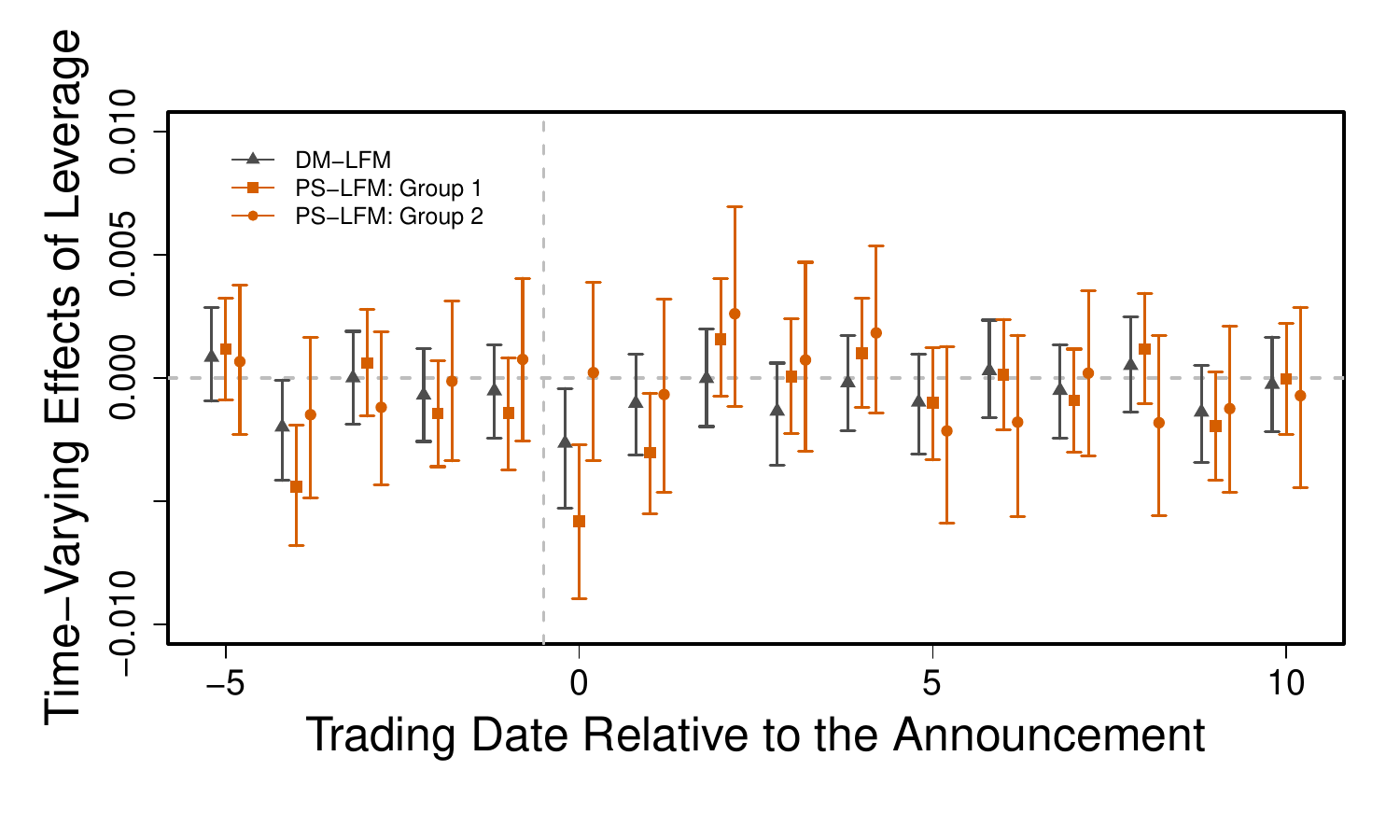} \\
    \end{tabular}
    \caption{Time-varying Effects of Firm-level Covaraites}\label{tvp}
\end{center}
{\footnotesize {\it Note}: The dots are the posterior means and the error bars represent corresponding  
95\% credible intervals.}
\end{figure}

We next examine whether the time-varying effects of firm-level covariates differ across propensity 
score strata and compare these estimates with those obtained from the DM-LFM. The results, shown in 
Figures~\ref{tvp}, indicate that the time-varying covariate effects can indeed 
vary across propensity score strata and, in some periods, differ substantially from the DM-LFM 
estimates.

As an additional robustness check, we conduct a placebo test by treating the two trading days prior 
to the announcement as placebo periods. The results, reported in Figures~\ref{datt_p} in 
Appendix~\ref{EA}, show that the 95\% credible intervals cover zero during the placebo periods for 
both model specifications, providing reassurance against spurious estimated abnormal returns.

\begin{figure}[!h]
\begin{center}
    \begin{tabular}{c}
         \includegraphics[scale = 0.5]{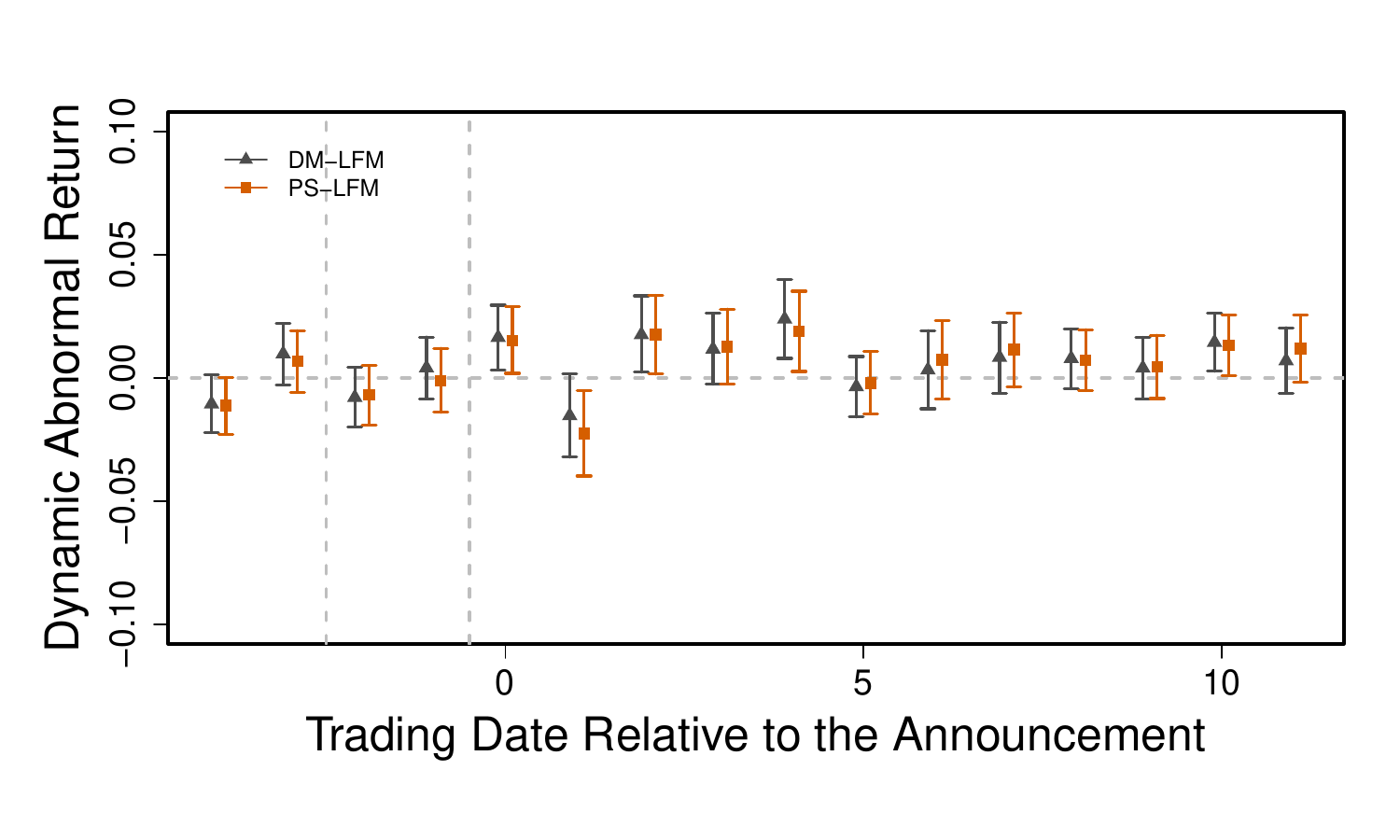}
    \end{tabular}
    \caption{Placebo Test on the Estimated Dynamic Effect}\label{datt_p}
\end{center}
{\footnotesize {\it Note}: The dots are the posterior means and the error bars represent corresponding  
95\% credible intervals.}
\end{figure}

\begin{figure}[!h]
\begin{center}
    \begin{tabular}{c}
         \includegraphics[scale = 0.45]{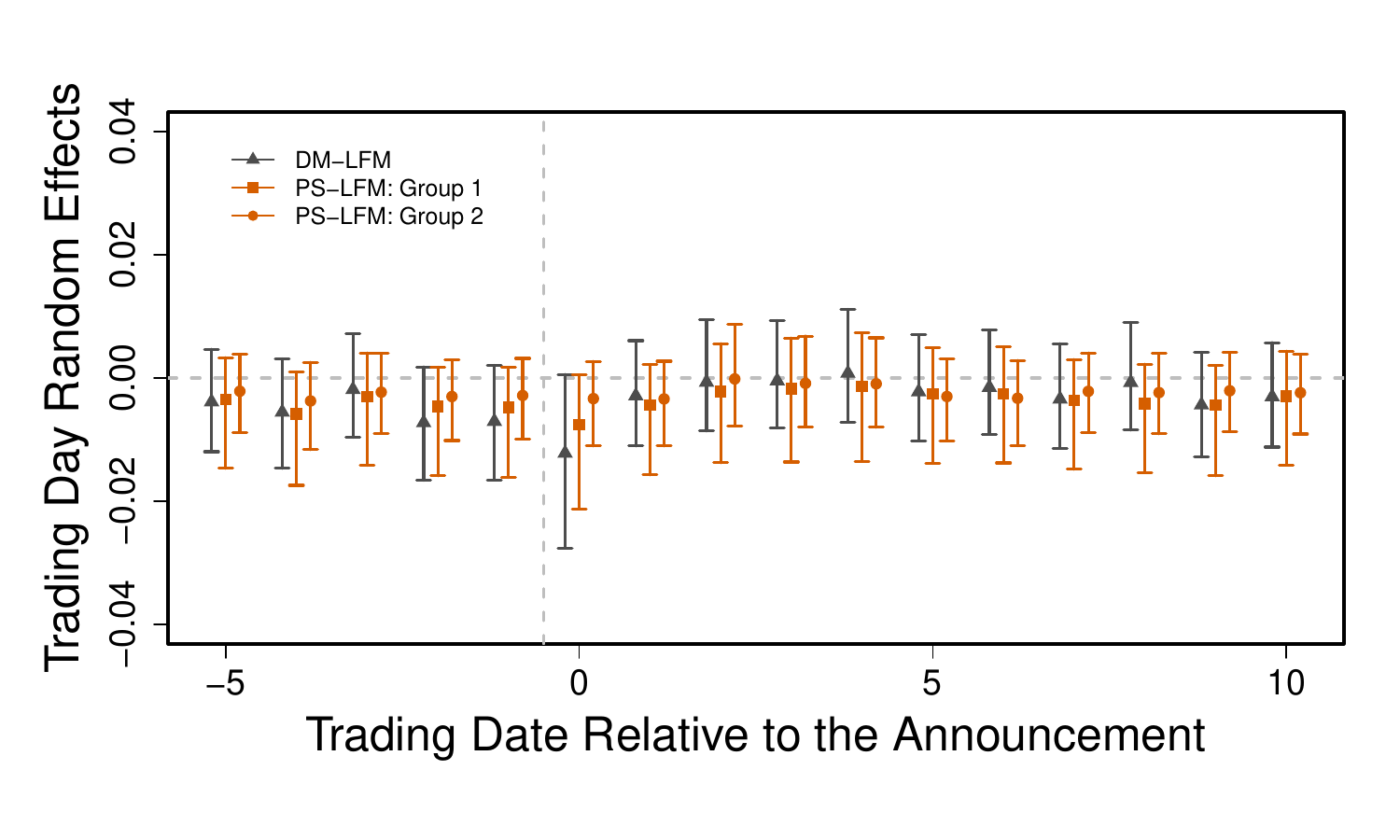} 
    \end{tabular}
    \caption{Estimated Time Random Effects}\label{tvp0}
\end{center}
{\footnotesize {\it Note}: The dots are the posterior means and the error bars represent corresponding  
95\% credible intervals.}
\end{figure}

\end{document}